\documentclass[12pt,preprint,dvips,deluxetable]{aastex}

  \newcommand{\etal}{et~al.\ }

\makeatletter
\newcommand\cutinheadb[1]{%
 \noalign{\vskip .8ex}%
 \@ptabularcr
 \noalign{\vskip -4ex}%
 \multicolumn{\pt@ncol}{c}{#1}%
 \@ptabularcr
 \noalign{\vskip .8ex}%
 \hline
 \@ptabularcr
 \noalign{\vskip -1.5ex}%
}%
\def\cutinheadb@ppt#1{%
 \noalign{\vskip .8ex}%
 \@ptabularcr
 \noalign{\vskip -1.5ex}
 \multicolumn{\pt@ncol}{c}{#1}%
 \@ptabularcr
 \noalign{\vskip .8ex}%
 \hline
 \@ptabularcr
 \noalign{\vskip -1.5ex}%
}%
\makeatother

\makeatletter
\newcommand\cutinheadc[1]{%
 \noalign{\vskip .8ex}%
 \@ptabularcr
 \noalign{\vskip -4ex}%
 \multicolumn{\pt@ncol}{c}{#1}%
 \@ptabularcr
 \noalign{\vskip .8ex}%
 \hline
 \@ptabularcr
 \noalign{\vskip -1.5ex}%
}%
\def\cutinheadc@ppt#1{%
 \noalign{\vskip .8ex}%
 \@ptabularcr
 \noalign{\vskip -1.5ex}
 \multicolumn{\pt@ncol}{c}{#1}%
 \@ptabularcr
 \noalign{\vskip .8ex}%
 \hline
 \@ptabularcr
 \noalign{\vskip -1.5ex}%
}%
\makeatother

\shorttitle{GLIMPSE Completeness Characterization}

\begin{document}

\slugcomment{Accepted for Publication in ApJS, 2013 May 28}

\title{EFFECTS OF DIFFUSE BACKGROUND EMISSION AND SOURCE CROWDING ON
PHOTOMETRIC COMPLETENESS IN
\emph{SPITZER SPACE TELESCOPE} IRAC SURVEYS: THE GLIMPSE CATALOGS AND ARCHIVES  }

\author{Henry A. Kobulnicky}
\affil{Department of Physics \& Astronomy \\  1000 E. University \\
University of Wyoming \\ Laramie, WY 82071 
\\ Electronic Mail: chipk@uwyo.edu}

\author{Brian L. Babler}
\affil{Department of Astronomy, University of Wisconsin-Madison,  \\
475 N. Charter St., \\ Madison, WI 53706
\\ Electronic Mail: brian@astro.wisc.edu}

\author{Michael J. Alexander}
\affil{Department of Physics \& Astronomy \\  1000 E. University \\
University of Wyoming \\ Laramie, WY 82071 
\\ Electronic Mail: malexan9@uwyo.edu}

\author{Marilyn R. Meade, Barbara A. Whitney, Edward B. Churchwell}
\affil{Department of Astronomy, University of Wisconsin-Madison,  \\
475 N. Charter St., \\ Madison, WI 53706
\\ Electronic Mail: meade@astro.wisc.edu,
bwhitney@astro.wisc.edu,ebc@astro.wisc.edu}


\begin{abstract}

We characterize the completeness of point source lists from
{\it Spitzer Space Telescope} surveys in the four Infrared
Array Camera (IRAC) bandpasses, emphasizing  the Galactic
Legacy Infrared Mid-Plane Survey Extraordinaire (GLIMPSE)
programs (GLIMPSE~I, II, 3D, 360; Deep GLIMPSE) and their
resulting point source Catalogs and Archives.   The analysis
separately addresses effects of  incompleteness resulting
from high diffuse background emission and incompleteness
resulting from point source confusion (i.e., crowding). An
artificial star addition and extraction analysis
demonstrates that completeness is strongly dependent on
local background brightness and structure, with
high-surface-brightness regions suffering up to five
magnitudes of reduced sensitivity to point sources.  This
effect is most pronounced at the IRAC 5.8 and 8.0 $\mu$m
bands where UV-excited PAH emission produces bright, complex
structures (photodissociation regions; PDRs).   With regard
to  diffuse background effects, we provide the completeness
as a function of stellar magnitude and diffuse background
level in graphical and tabular formats.   These data are
suitable for estimating completeness in the
low-source-density limit in any of the four IRAC bands in
GLIMPSE Catalogs and Archives and some other $Spitzer$ IRAC
programs that employ similar observational strategies and
are processed by the GLIMPSE  pipeline.  By performing the
same analysis on smoothed images  we show that the  point
source incompleteness is primarily a consequence of
\emph{structure} in the diffuse background emission rather
than photon noise.  With regard to source confusion in the
high-source-density regions of the Galactic Plane, we
provide figures illustrating the 90\% completeness levels as
a function of point source density at each band.   We
caution that completeness of the GLIMPSE~360/Deep GLIMPSE
Catalogs is suppressed relative to the corresponding
Archives  as a consequence of rejecting stars that lie in
the PSF wings of  saturated sources.  This effect is minor
in regions of low saturated star density, such as toward the
Outer Galaxy;  this effect is significant along sightlines
having a high density of saturated sources,  especially for
Deep GLIMPSE and other programs observing closer to the
Galactic center using 12 s or longer exposure times.

\end{abstract}


\section{Introduction } 

The {\it Spitzer Space Telescope} \citep{Werner04} has
conducted numerous wide-area surveys, yielding a rich legacy
of image data and point source catalogs that have become
heavily used commodities.   In terms of sky coverage, the
largest of these are the Galactic Legacy Infrared Mid-Plane
Survey Extraordinaire programs \citep[GLIMPSE, GLIMPSE~II,
GLIMPSE~3D;][]{benjamin,churchwell2009}  covering almost 400
square degrees of the Milky Way Plane in all four 
mid-infrared bands with the Infrared Array Camera
\citep[IRAC; ][]{Fazio04}.  The $Spitzer$ Mapping of the
Outer Galaxy \citep[SMOG;][]{careysmog}, the $Spitzer$
Legacy Survey of the Cygnus-X Complex \citep[][]{hora}, and
the Vela-Carina large program \citep{zasowski09}  together
cover another 120 square degrees.   The warm $Spitzer$
mission program GLIMPSE~360 \citep{Whitney08} imaged
$\sim$511 square degrees of the outer Galactic Plane in the
3.6 and 4.5 $\mu$m bands.  Deep GLIMPSE \citep{Whitney11} 
re-imaged a similar portion of the Galactic Plane  as
GLIMPSE I/II, but only at 3.6 and 4.5 $\mu$m, using a longer
10.4 s integration time.  Other Galactic programs and
numerous extragalactic legacy programs provided much deeper
datasets but over relatively small fields of view. Our focus
here is primarily the Galactic science programs that have
yielded large lists of point sources.\footnote{See
http://irsa.ipac.caltech.edu/data/SPITZER/docs/spitzermission/observingprograms/
for a summary of {\it Spitzer} Legacy and other large
programs. Data product and photometry documents for the
GLIMPSE programs may be found at
http://www.astro.wisc.edu/glimpse/docs.html and
http://irsa.ipac.caltech.edu/data/SPITZER/docs/spitzermission/observingprograms/legacy/glimpse/} 
Table~\ref{surveys.tab} summarizes the approximate
coverages, exposure times, observational bandpasses, and
total number of  photometered point sources for these and
other Galactic {\it Spitzer} survey.  There are two types of
GLIMPSE source lists:  a high reliability point source
``Catalog'' and a more complete point source ``Archive''. 
The source lists are a result of doing photometry on each
IRAC frame, averaging all detections made in a single band
(in-band merge), then merging the photometry from all
wavelengths, including 2MASS $JHK_s$ sources, at a given
position on the sky (cross-band merge).  The GLIMPSE source
list criteria have been developed to ensure that each source
is a legitimate detection and that the fluxes reported for
the IRAC bands are of high quality. As of this writing,
these $Spitzer$ programs have generated nearly  158 million
sources in the highly reliable  \emph{GLIMPSE Point Source
Catalogs} (GPSC) and 229 million in the more complete
\emph{GLIMPSE Point Source Archives}  (GPSA).  

The utility of point source catalogs resulting from the
{\it Spitzer} mission depends, partly, on understanding their
completeness (i.e., the ratio of true sources detected to
the number of all true sources).     The stellar magnitude
at which the GPSC or GPSA is nominally complete varies with
bandpass.  It also varies greatly as a function of
background surface brightness and as a function of local
source density where confusion with neighboring sources
limits detection and photometry.  Figure 2 of \citet{robit08} 
shows how high diffuse background levels inhibit
the detection of point sources in the {\it Spitzer} 8.0 $\mu$m band,
which is affected by strong, diffuse emission features
arising from polycyclic aromatic hydrocarbons (PAHs) in and
around regions of star formation.  The IRAC 5.8 $\mu$m band,
and to a much lesser extent the 3.6 $\mu$m band, also
encompass these features, leading to strong spatial
completeness variations, especially in complex fields near
the Galactic Plane. It is widely recognized that sensitivity
to point sources is significantly reduced in high-background
fields and fields having high  point  source densities, but
there has not yet been a general quantitative analysis of
this effect.  

Our goal in this contribution is to quantify the
photometric completeness in the four {\it Spitzer} IRAC
bandpasses, considering separately the
effects of 1) high diffuse backgrounds, and 2) high point
source surface densities.  In Section 2 we describe  an artificial
star analysis procedure for addressing the effects of
incompleteness resulting from diffuse background emission.  
In Section 3 we provide plots and tables of completeness as
a function of diffuse background level and stellar
magnitude.  In Section 4 we address the effects of
incompleteness resulting from source crowding (i.e.,
confusion), and we provide plots and tables of the magnitude
at which 90\% completeness  is achieved as a function of
point source surface density.  All magnitudes referenced
herein refer to the Vega magnitude system used by the 
GLIMPSE pipeline and data products.

GLIMPSE~I obtained two 1.2 s exposures of each object, while
GLIMPSE~II obtained three such 1.2 s exposures.   GLIMPSE~3D
utilized either two or three 1.2 s exposures at each
location. GLIMPSE~360 and Deep GLIMPSE employed the
high-dynamic-range (HDR) mode to obtain  three 0.4 and 10.4
s exposures at each position.  We adopt the GLIMPSE~I/II/3D
survey strategy and its resulting GPSC and GPSA as our
baseline dataset, and we perform a similar analysis for the 
GLIMPSE~360/Deep~GLIMPSE survey strategy.    Although it is
unrealistic to perform a completeness characterization for
every possible survey strategy and sightline, we intend that
these two results be general enough that consumers of 
source lists generated as part of GLIMPSE and similarly
processed {\it Spitzer} IRAC surveys, such as SMOG, Cygnus-X, and
Vela-Carina can infer completeness for those programs as
well.  This completeness analysis is pertinent only to 
point sources and is not appropriate for extended sources.

\section{Artificial Star Analysis Procedure for Diffuse Background Completeness }

\begin{figure} \plotone{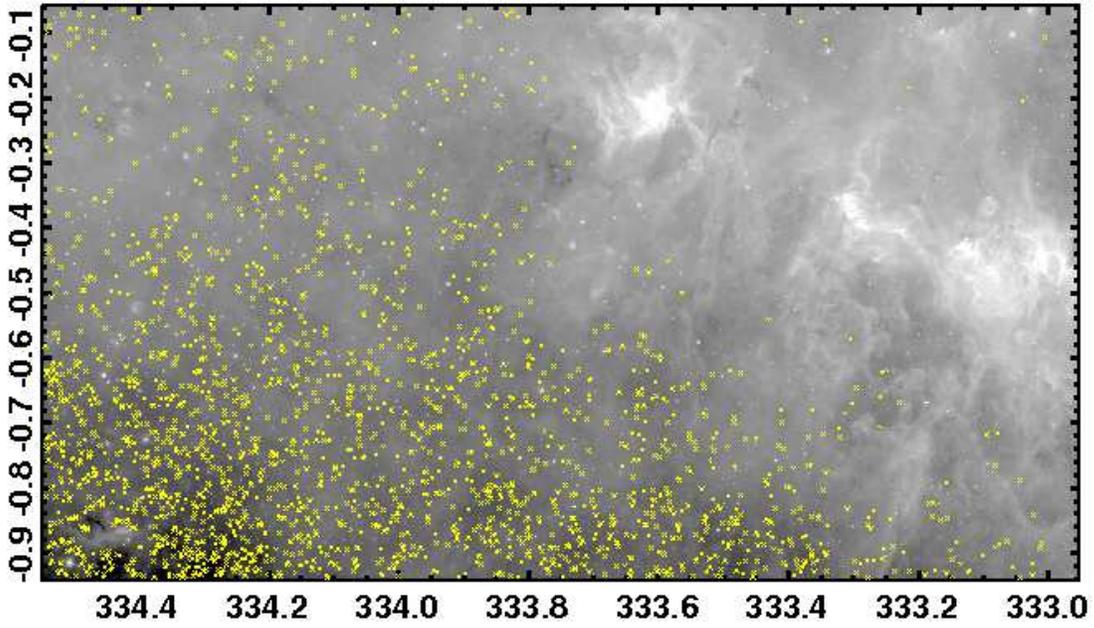} \caption{Logarithmic greyscale
representation of the IRAC 8.0 $\mu$m mosaic from the  GLIMPSE~I
survey near $\ell$=334\degr.  Crosses  mark detected point sources
fainter than 12th magnitude from the GLIMPSE Point Source Catalog. 
The variation in source density reflects the varying completeness
levels as a function of diffuse background emission.       
\label{pretty} } \end{figure}

We selected a $\sim$1.5 square degree region of the GLIMPSE
I survey between $333.0\degr<\ell<334.5\degr$, $-1\degr<b<0\degr$,
exhibiting a large range in diffuse background level to use
as a test region. This area contains some of the brightest
mid-IR background regions in the Galactic Plane. 
Figure~\ref{pretty} shows a logarithmic greyscale
representation of this region in the 8 $\mu$m IRAC band.  
This region of the Galactic Plane spans a
large dynamic range from 20 MJy/sr in the lower left to over
5000 MJy/sr.  Yellow crosses show detected 8.0 $\mu$m point
sources fainter than 12th magnitude as included in the
GLIMPSE Point Source Catalog.  The varying density of
cataloged sources from lower left to upper right 
dramatically  illustrates the effects of high and complex
background levels on completeness.   

The GLIMPSE~I data in this region is comprised of $\sim$600
individual 1.2-second exposures in each of the four $IRAC$
bandpasses.  The IRAC pixel size is 1.2\arcsec\ at all four
bandpasses. To each 256$\times$256 pixel basic calibrated
data (BCD) image we added roughly 600 artificial stars at
random locations using an average PSF appropriate for each
band, as constructed by the GLIMPSE data processing team
from an ensemble of IRAC point sources.   The PSF was scaled
to a random magnitude between $M_{low}$ and $M_{up}$,
Gaussian noise ($\sqrt{N_{photons}}$) was added to each
artificial star PSF on a pixel-by-pixel basis, and the
artificial star was added at a (pseudo-) random location  to
the BCD image. The center locations of the artificial stars
were constrained to 1) lie no closer than 5 pixels from the
edge of the image, and 2) lie no closer than 3 pixels to a
real star detected during previous processing by the GLIMPSE
pipeline (dubbed the ``Actual'' list of point sources). 
This avoids both edge effects and the effects of confusion
from neighboring point sources in order to focus on effects
solely related to diffuse background levels.  We set
$M_{low}$=8 $M_{up}$=15 for all bands, spanning seven
magnitudes in brightness to cover nearly the entire dynamic
range of the $GLIMPSE$ survey in order to sample both very
complete and very incomplete flux levels. The artificial
stars have the same magnitude in each band.  Effectively,
this means that  we are simulating sources having zero
color, which comprise the vast majority ($\sim$94\% of
sources have  $| m_2 - m_1 | < 0.5$ ) of all GPSC sources
having  measurements in all four bands.   \footnote{This
means that completeness for stars having unusual colors
(e.g., very red objects such as young stellar objects or
heavily reddened stars) is not accurately reported by our
metric ``Completeness1'' (defined below), but completeness
of such red sources is tabulated meaningfully by the
``Completeness2'' metric. } However,  it is not practical to
conduct a general analysis for arbitrary  spectral energy
distributions.  The GLIMPSE pipeline was run to generate a
list of point sources from the Actual+Artificial images in
the same manner as the published \emph{GLIMPSE Point Source
Catalogs} and \emph{GLIMPSE Point Source Archives}.  

Inclusion in the highly reliable \emph{GLIMPSE Point Source
Catalog}  requires that a source be
detected\footnote{Details of the detection algorithm
implemented in the GLIMPSE pipeline appear in the Appendix.}
{\it twice} (i.e., in at least two individual IRAC
exposures; detection is performed on the  BCD frames) in any
one IRAC band at S/N$>$5 and {\it at least once} in any
adjacent band at S/N$>$5, where the 2MASS K$_s$ band is
allowed as an adjacent band for IRAC 3.6 $\mu$m.     Hence,
it is possible for a source to be included in the GPSC but
lack photometry in as many as two (or even three) of the
four IRAC  bands.  Most commonly, non-detections occur at
IRAC bands 5.8 and/or  8.0 $\mu$m as a result of
the reduced instrumental sensitivity and the decreasing flux
on the Rayleigh-Jeans tail of the spectral energy
distribution for the majority of astrophysical sources. 
Objects with rising spectral energy distributions, such as
protostars, evolved stars exhibiting circumstellar excesses,
or heavily extincted objects, could have detections at IRAC
5.8 \& 8.0 $\mu$m but not at the shorter wavelengths. 
Inclusion in the Catalog also requires that 1) a source does
not fall within the PSF wings of a saturated source (as
defined in the GLIMPSE data release documentation), 2) no
hot or dead pixel lies within 3 pixels  of source center, 3)
the source is not within 3 pixels of a frame edge, and 4)  
no Archive source lies within  2\arcsec\ of the source.
Furthermore, the sources used to identify and confirm a
detection  must be brighter than 0.6 mJy (14.2 mag), 0.4 mJy
(14.1), 2 mJy (11.9), and 10.0 mJy (9.5) in the 3.6, 4.5,
5.8, and 8.0 $\mu$m bands, respectively. However, sources
fainter than these limits at a particular band may appear in
the Catalogs  if they are identified and confirmed using
\emph{other}  bands. Users should consult the GLIMPSE data
release documentation\footnote{ 
http://irsa.ipac.caltech.edu/data/SPITZER/GLIMPSE/doc/glimpse1\_dataprod\_v2.0.pdf}
for a full description of source selection and rejection
criteria.  

Inclusion in the less reliable but more complete
\emph{GLIMPSE Point Source Archive} requires  that a source
be detected at least once in any two bands at S/N$>$3 and
that there are no neighboring sources within 0.5\arcsec.  
Sources that lie within the PSF wings of a saturated star
\emph{are} included in the Archive.  This last criterion
means that the completeness we estimate for the Catalog is
going to be suppressed  relative to the Archive at all
magnitudes and at all background levels because of the
presence of saturated stars.  The saturation limit for the
short 1.2 s GLIMPSE~I/II/3D frame times is approximately  7,
6.5, 4, and 4 magnitudes (439, 450, 2930, and 1590 mJy)  for
IRAC bands 3.6, 4.5, 5.8, and 8.0 $\mu$m, respectively. The saturated source areal
density is a strong function of Galactic  position but is on
the order of $\sim$40 per square degree at $\ell=36$\degr,
as estimated from 2MASS K-band source counts.\footnote{The
2MASS K$_S$ band is used as a proxy for the IRAC 3.6 $\mu$m
band because the GLIMPSE pipeline does not detect or extract
saturated sources.}    Therefore, the effect of saturated
stars on completeness in the GLIMPSE~I/II/3D Catalogs and
similar shallow survey Catalogs is minimal, given that the
PSF wings ($R<$24 pixels, but an irregular region given by
the shape of diffraction spikes) comprise $\lesssim$1\% of
the survey area.    

To characterize the completeness at each band the list of
detected point sources from the ``Actual+Artificial'' images
was compared to the input list of artificial sources (i.e.,
the Artificial list). Any detected source within 1.2 pixels
of a source in the ``Artificial'' list was considered to be
a successfully recovered  source. We  find that the
magnitudes of recovered stars always match the magnitudes of
inserted stars within the expected uncertainties, typically
0.05 mag at the bright end and $\sim$0.5 mag at the faint
end. Therefore, the matching process does not include a
magnitude  criterion. We found it necessary to develop two
definitions of completeness. In the first definition, dubbed
Completeness1, we define completeness as the  number of
artificial sources present in the resulting GPSC divided by
the  number of input artificial sources.  This inclusive
definition allows for the possibility that a given star may
be included in the GPSC {\it even though it may lack a
detection} at any given band, N.  In the second definition,
Completeness2, we require that a given source actually be
detected in band N specifically, rather than being included
in the Catalog only by virtue of its detection at other
bands.  The more restrictive Completeness2
values are always lower than Completeness1, especially at
5.8 and 8.0 $\mu$m where the deleterious effects of high and
structured backgrounds on source identification are most
severe.

\section{Diffuse Background Completeness Results }  
\subsection{GLIMPSE I, II, 3D Point Source Catalog (and Vela-Carina)}

\begin{figure} \plotone{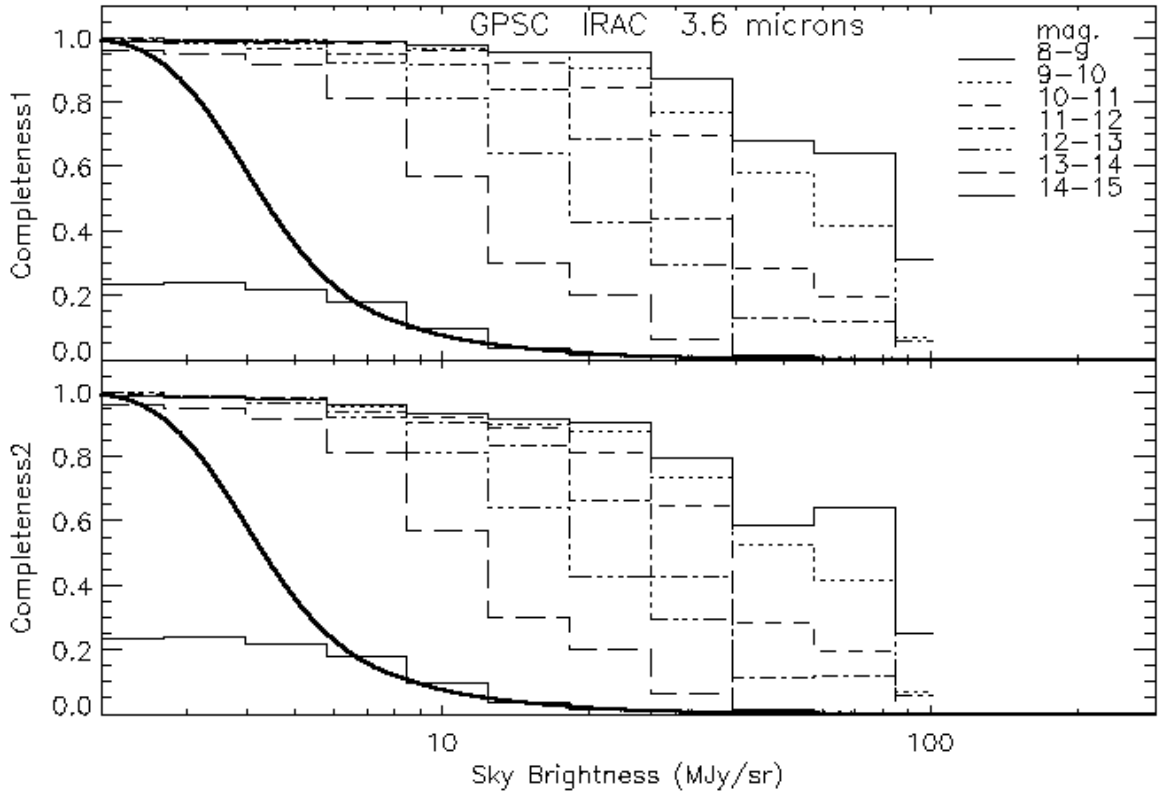} \caption{GLIMPSE I/II/3D Catalog 
Completeness1 (upper panel) and Completeness2 (lower panel)
versus sky brightness for IRAC 3.6 $\mu$m as a function of
stellar magnitude, denoted by the line style from 8th to
15th magnitude. The heavy solid curve shows the cumulative 
fraction of sky background regions brighter than a given level.  For example,
approximately 35\% of the background regions in our test
area are brighter than 5 MJy sr$^{-1}$ at 3.6 $\mu$m.   
\label{comp1a} } 
\end{figure}

Figure~\ref{comp1a} (upper panel) plots Completeness1 versus
background sky brightness in MJy sr$^{-1}$ for the IRAC 3.6
$\mu$m artificial star simulations.  The ordinate is
partitioned into logarithmically spaced sky brightness bins
to increase the reliability of high-surface-brightness bins 
which comprise only a small fraction of the test area. 
Linestyles indicate stellar magnitude bins from 8th to 15th
magnitude, as shown by the legend. The heavy solid curve
shows the cumulative distribution of the background sky
brightness. For example, the solid curve shows that
approximately  35\% of the background regions in our test
area are brighter than 5 MJy sr$^{-1}$ at 3.6 $\mu$m. The
rapid decline in completeness at background levels greater
than about 10 MJy~sr$^{-1}$  is clear in all but the
brightest magnitude ranges.   Furthermore, the faintest
magnitude bin, 14th -- 15th magnitude, is never more than
25\% complete even in the  regions of lowest sky
brightness.   We attempted to fit simple analytical formulae
to these data to provide an easily integrable functional
form for completeness as a function of magnitude and
background level, but no simple form (Gaussian, exponential,
trigonometric, polynomial) yielded good fits across the
range of magnitudes and background levels pictured in
Figure~\ref{comp1a}. We  therefore elected to simply
tabulate the results and allow end users to perform their
own interpolation or fitting. Table~\ref{comp1cat.tab} lists
numerical values for Completeness1 as a function of
magnitude and background level for the 3.6 $\mu$m bandpass
as shown graphically in Figure~\ref{comp1a}.  Note that the
uncertainties on the tabulated completeness are at the level
of 10--15\% for  the bin at the highest background levels as
a result of statistical limitations, but in the other bins
uncertainties are typically $\leq$2\%.   Statistics in the
lowest surface brightness bin are also uncertain at the
level of a few percent as a result of the paucity of
extremely dark regions in the Galactic Plane.

Figure~\ref{comp1a} (lower panel) plots Completeness2 versus
background sky brightness for the IRAC 3.6 $\mu$m. 
Table~\ref{comp2cat.tab} lists these results in the same
manner as for Completeness1 in Table~\ref{comp1cat.tab}. 
The results are nearly identical to the  Completeness1 case
because PAH emission from diffuse background  sources is 
relatively minor at 3.6 $\mu$m and the flux from stellar
photospheres is a maximum, thereby aiding source
identification and extraction.   Sources in the GPSC are
most commonly included by virtue of having a detection at
IRAC 3.6 $\mu$m.

\begin{figure} \plotone{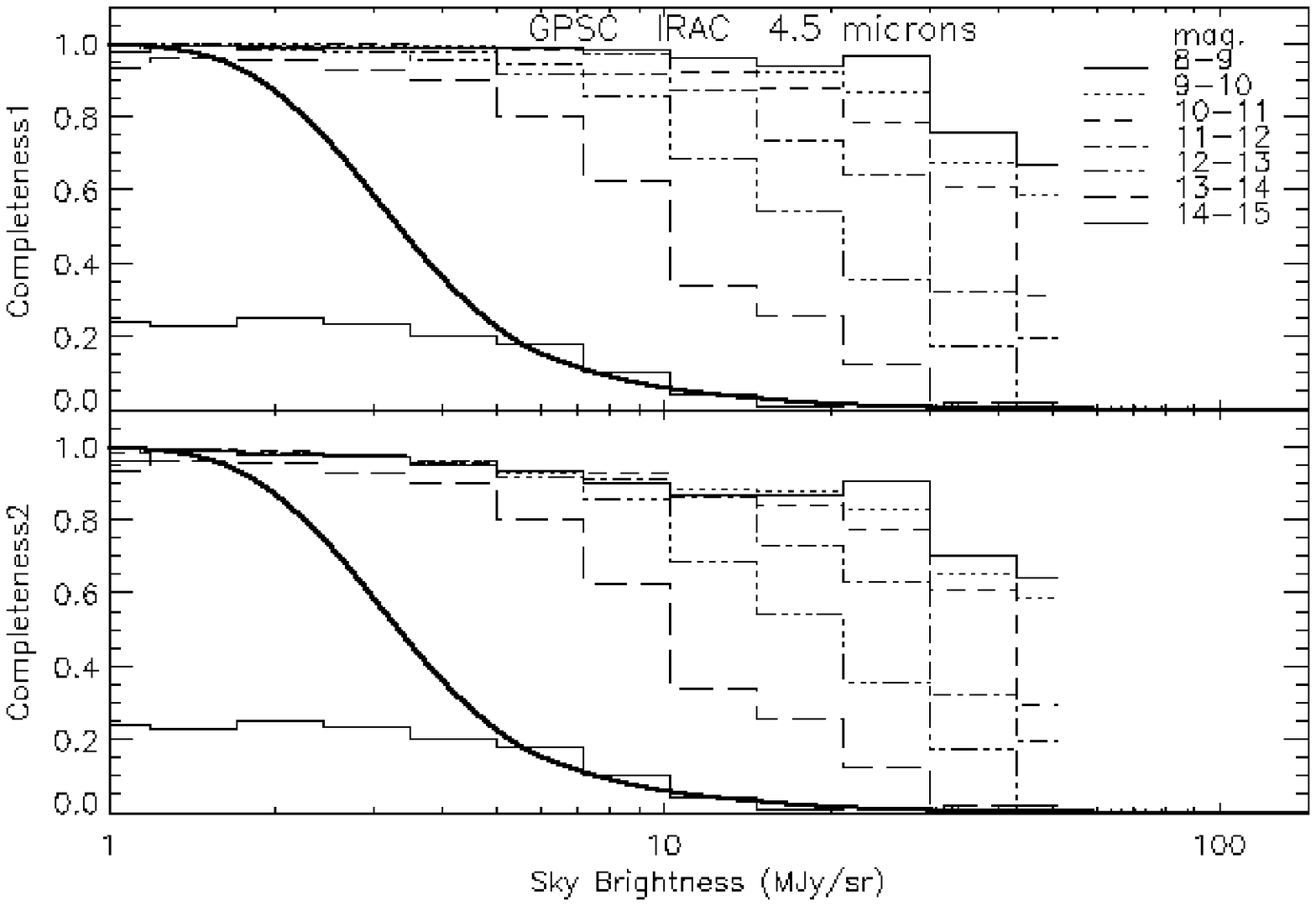} \caption{GLIMPSE I/II/3D Catalog 
Completeness1 (upper panel) and Completeness2 (lower panel)
versus sky brightness for IRAC 4.5 $\mu$m as a function of
stellar magnitude, with symbols as in Figure~\ref{comp1a}.      
\label{comp2a} } 
\end{figure}

Figure~\ref{comp2a}  plots Completeness1 and Completeness2
versus background sky brightness for the IRAC 4.5 $\mu$m
artificial star  simulations.  The overall trends are
similar to 3.6 $\mu$m where both Completeness1 and
Completeness2 decline markedly  in backgrounds above about
10 MJy sr$^{-1}$.  Only the brightest stars, $<$11th
magnitude, are readily detected and show a slower drop in
completeness.  The numerical results for Completeness2 are
nearly identical to Completeness1.   This results from the
fact that background levels are low in the 4.5 $\mu$m band
(no PAH features) and that sources in the GPSC are almost
always included by virtue of having a detection at 4.5
$\mu$m.   Tables~\ref{comp1cat.tab} and \ref{comp2cat.tab}
record these numerical completeness data for the 4.5 $\mu$m
analysis.

\begin{figure} \plotone{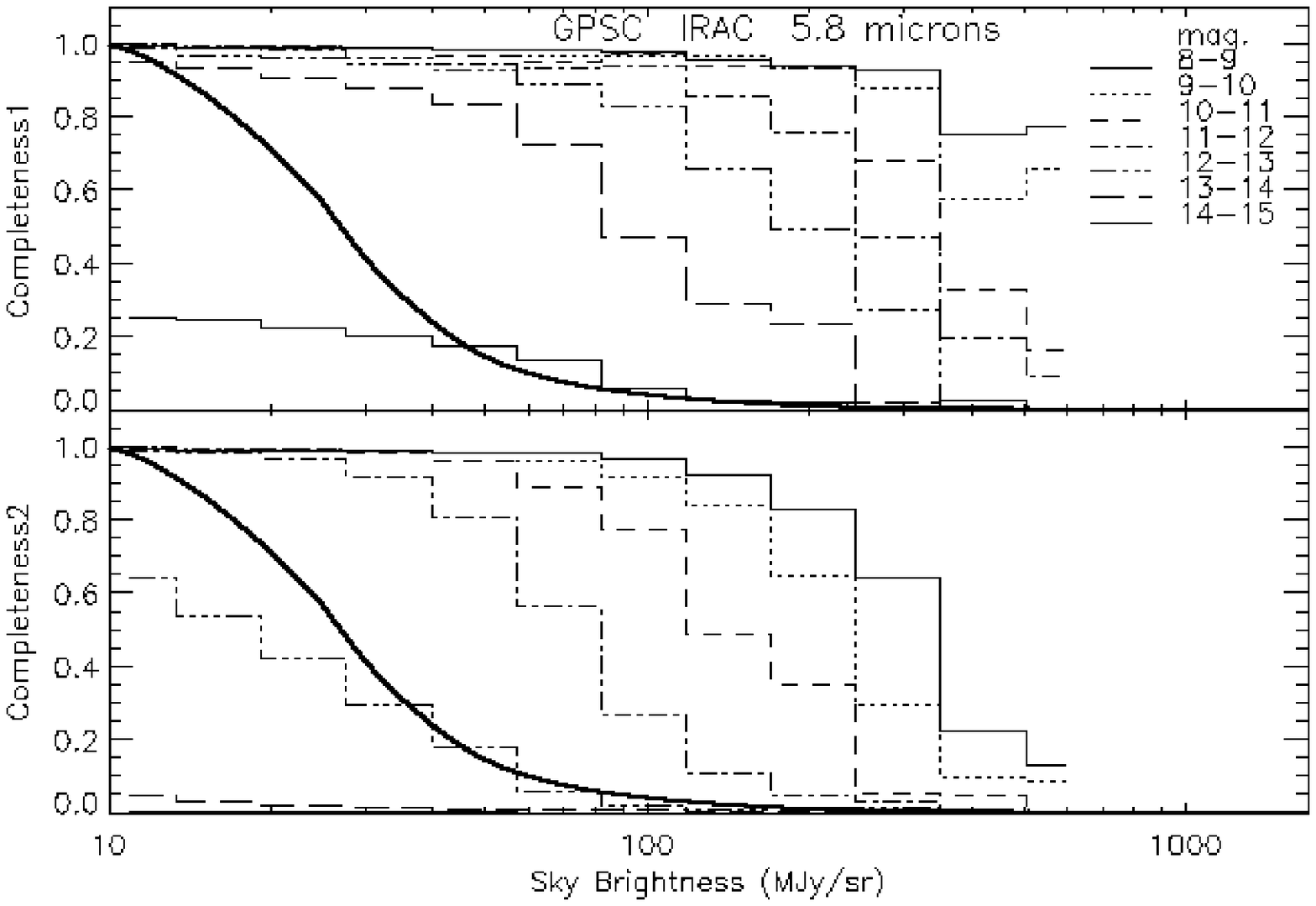} \caption{GLIMPSE I/II/3D Catalog 
Completeness1 (upper panel) and Completeness2 (lower panel)
versus sky brightness for IRAC 5.8 $\mu$m as a function of
stellar magnitude, with symbols as in Figure~\ref{comp1a}.      
\label{comp3a} } 
\end{figure}

Figure~\ref{comp3a}  plots Completeness1 and Completeness2
versus background sky brightness for the IRAC 5.8 $\mu$m
simulations.   These histograms show dramatically decreased
completeness levels for all but the brightest stars as the
background levels exceed  $\sim$50 MJy sr$^{-1}$.  The IRAC
5.8 $\mu$m band contains strong PAH features, and regions
with specific intensities exceeding 200 MJy sr$^{-1}$ are
not uncommon in this and other major star-forming complexes.
A comparison of  Figure~\ref{comp3a} shows that 
Completeness2 is significantly lower than Completeness1 for
stars fainter than about 11th magnitude for all background
levels greater than $\sim$100 MJy sr$^{-1}$.  This results
from sources being included in the Catalog (Completeness1)
but not having a detection at 5.8 $\mu$m (Completeness2).  
Tables~\ref{comp1cat.tab} and \ref{comp2cat.tab} record
these numerical completeness data for the 5.8 $\mu$m
analysis.

Figure~\ref{comp4a}  plots Completeness1 and Completeness2
versus background sky brightness for the 8.0 $\mu$m
simulations. While the behavior of the calculated
completeness is similar to the plots for 5.8 $\mu$m above,
the difference  between Completeness1 and Completeness2 is
even more dramatic.   Completeness2 shows a much more rapid
decline than Completeness1, to the point where sources
fainter than 12th magnitude are rarely detected in any but
the darkest background regions.   The difference between
the upper and lower panels of Figure~\ref{comp4a}  illustrates why the number of 8.0
$\mu$m detections in GLIMPSE averages $<$30\% the number of
3.6 $\mu$m detections for a typical low-latitude sightline.
In regions with background levels greater than 100 MJy
sr$^{-1}$, the lower panel shows that only 11th magnitude
and brighter sources are likely  to be represented in the
Catalog and have a measurement at 8.0 $\mu$m.  However, on a
more  positive note, the upper panel shows that faint 8.0
$\mu$m sources \emph{are} likely to be represented in the
Catalogs by virtue of having measurements at other bands,
principally at 3.6 and 4.5 $\mu$m.  

In summary, Figures~\ref{comp1a} through \ref{comp4a}, in
conjunction with Tables~\ref{comp1cat.tab} and
\ref{comp2cat.tab} provide a means of estimating the likelihood
of detecting sources in a particular band and being included  in
the GLIMPSE Point Source Catalogs.  These Figures and Tables
should also be applicable to other {\it Spitzer} IRAC
observations using similar short exposure times, low
point source densities, and a small
number of observations  of each target, such as the
Vela-Carina project which was processed by the GLIMPSE pipeline. 

\subsection{GLIMPSE I, II, 3D Point Source Archive}

We performed the same analysis for the GPSA as for the GPSC.
Figures~\ref{comp1aA} through \ref{comp4aA} plot
Completeness1 and Completeness2 as a function of background
brightness for the Archive.  Tables~\ref{comp1arc.tab} and
\ref{comp2arc.tab} summarize these results numerically.  A
comparison of Figures~\ref{comp1aA} through \ref{comp4aA}
with  the corresponding sequence in Figures~\ref{comp1a}
through \ref{comp4a} reveals that more stars are recovered
in the Archives versus the Catalogs at all bands.  The
typical difference is a few percent to as much as a few tens
of percent,  with the most dramatic improvement being at low
background levels for the faintest stars.  

\subsection{GLIMPSE 360 \& Deep GLIMPSE}

GLIMPSE 360 and Deep GLIMPSE employed the IRAC
high-dynamic-range mode using 3.6 and 4.5 $\mu$m bands during the
{\it Spitzer} warm mission to obtain three 0.4 and three
10.4-second exposures at each location along a large portion
of the Galactic Plane  following the warp of the disk in the
outer Milky Way.    We employed the same artificial star
methodology described above in order to generate
completeness data for the 3.6 and 4.5 $\mu$m photometry
obtained in the course of these large programs.  The
selected test region covers a wide range of background
levels in the region $133.5\degr<\ell<135.0\degr$ and
$0.0\degr<b<1.9\degr$ obtained during the GLIMPSE 360
observing program.  The GLIMPSE~360 completeness analysis 
adds and recovers artificial stars only on the 10.4 s exposures, 
since we are interested primarily in  completeness for the
faint end of the observed magnitude range.   

The GLIMPSE 360 and Deep GLIMPSE observing strategy is
sufficiently different from the preceding GLIMPSE programs
that slightly different criteria are applied to determine
inclusion in the Catalogs and Archives.   Given three
HDR sequences (an HDR sequence means one 0.4 s exposure plus one
10.4 s exposure),  there are six possible frames at each
location in each of two bands  where a source might be
detected.   In general, M is the number of detections for a
given source  out of a possible N detections. Inclusion in
the GLIMPSE~360/Deep GLIMPSE Catalog requires M/N$>$0.6 in
one band (the selection band) and M/N$>0.32$ in an adjacent
band (the confirmation band), with S/N$>$5 for both bands.  
Detection in the 2MASS K$_S$ band counts as a confirmation,
but in practice, only a few percent of all GLIMPSE~360
sources are included in the Catalog by virtue of having a
K-band detection in addition to a 3.6 $\mu$m detection.   
Additionally, inclusion in the Catalog requires that 1) the
source be fainter than the non-linear regime for the IRAC
array (10.25 and 9.6 mag at 3.6 and 4.5
$\mu$m, respectively, for the IRAC 10.4 s exposure times; the HDR 0.4 s data 
has brighter limits but is not considered here in our simulations), 
2) that the source not lie within
2\arcsec\ of an Archive source, 3) that the source not lie
within the  diffraction spikes of a saturated source, and 4)
that no hot or dead pixels lie within 3 pixels of the source
center.  Inclusion in the Archive requires the same M/N
limits but  sources within 0.5\arcsec\ of another source are
allowed, as are sources within the PSF wings of a saturated
source.   End users should consult the GLIMPSE~360 Data
Delivery 
Document\footnote{http://www.astro.wisc.edu/glimpse/glimpse360\_dataprod\_v1.1.pdf} 
for a full understanding of source flags and subtle criteria
that define the Catalog and the Archive sources.

Figure~\ref{comp5a}  shows the Completeness1 and 
Completeness2 results for the GLIMPSE 360/Deep GLIMPSE
Catalog for the 3.6 $\mu$m band.  Figure~\ref{comp6a}  shows
the Completeness1 and  Completeness2 Catalog results for 4.5
$\mu$m.  The histogram linestyles indicate magnitude ranges
from 11 -- 18.   Tables~\ref{G360C1.tab} and
\ref{G360C2.tab} report numerical data for the
GLIMPSE~360/Deep GLIMPSE Catalog simulations for the 3.6 and
4.5 $\mu$m bands. These figures show that a similar trend to
the 3.6 and 4.5 $\mu$m results from the GLIMPSE I/II/3D
simulations reported above.   Completeness varies as a
function of both magnitude and background level. Given the
deeper exposures for GLIMPSE 360/Deep GLIMPSE compared to
GLIMPSE I/II/3D,  the completeness is generally greater at
any given magnitude at any given background level.  One
noteworthy feature of  these histograms is the drop in
completeness at all magnitudes, even at relatively faint sky
background levels.  This means that the GPSC for GLIMPSE
360/Deep GLIMPSE is almost never 100\% complete, even at the
brightest magnitudes and darkest sky backgrounds. The reason
for this is described subsequently.    

Figure~\ref{comp7a} shows the Completeness1 and 
Completeness2 results for the GLIMPSE 360/Deep GLIMPSE
Archive for the 3.6 $\mu$m band.  Figure~\ref{comp8a}  shows
the Completeness1 and  Completeness2 results at 4.5 $\mu$m. 
Tables~\ref{G360A1.tab} and \ref{G360A2.tab} report
numerical data for the GLIMPSE~360/Deep GLIMPSE Archive
simulations for the 3.6 and 4.5 $\mu$m bands. These figures
and tables show that the Archive Completeness1 is very
similar to Completeness2.   This is expected, given that the
lack of IRAC 5.8 and 8.0 $\mu$m data in GLIMPSE~360 and
similar surveys essentially means that all of the sources
must be detected both at 3.6 and 4.5 $\mu$m in order to
appear in the Catalog or the Archive.  

Notably, GLIMPSE~360 Archive completeness in
Figures~\ref{comp7a} through  \ref{comp8a} is higher than
Catalog completeness in Figures~\ref{comp5a} through
\ref{comp6a}, \emph{even at bright magnitudes and low
background levels}. This results from the inclusion of
sources in the Archive  that lie within the  PSF wings of
saturated stars.  Such sources are rejected from the Catalog
in order to ensure high reliability.   Saturated stars are a
much greater problem in the longer exposures used for
GLIMPSE~360/Deep GLIMPSE compared to GLIMPSE I/II/3D.   
This means the photometry for a significant fraction of the areal coverage in
GLIMPSE 360/Deep GLIMPSE Catalogs is nullified because it lies 
under the PSF wings of saturated stars.
 The GLIMPSE~360/Deep GLIMPSE Archives, by contrast,
are significantly more complete because they include sources
that lie in the PSF wings of saturated stars.  The magnitude
of this effect in the GLIMPSE~360/Deep GLIMPSE  Catalogs 
varies dramatically with Galactic positions, being on the
order of a few percent in the outer Galaxy but rises to many
tens of percent in crowded regions of the inner Galaxy. 
Figure~\ref{saturate180} and  Figure~\ref{saturate030}  
show greyscale representations of the GLIMPSE~360/Deep
GLIMPSE  IRAC 3.6 $\mu$m mosaic near $\ell=180$\degr\ and
$\ell=030$\degr, respectively.  Red circles in the left
panel of each figure  mark sources included in the GPSC,
while green circles in the right panels mark sources
included in the GPSA. The effect of ``dead zones'' around
bright stars is dramatic in the left panel of
Figure~\ref{saturate030} where high stellar densities 
cause the removal of photometric results from the GPSC 
for a significant fraction of the
survey area.  By contrast, the effects of
saturated sources in the GPSC in Figure~\ref{saturate180}
for the $\ell=180$\degr\ field are rather subtle.  Here and
in other such fields having low stellar densities, the
completeness of the Catalog and Archive will be more similar
than for high-density fields, as only a few percent of the
survey area  falls under the wings of saturated stars.    
In summary, Catalog incompleteness in GLIMPSE~360/Deep
GLIMPSE varies greatly with location, and users are
encouraged to consider the merits of the GPSA where high
levels  of completeness are desired.

\section{Role of Structured Backgrounds }

Having empirically assessed the completeness simply as a
function of one parameter, i.e., background level, we now
consider the role that the structure of the diffuse
background emission plays in limiting point source  source
detection.  Figure~\ref{pretty} shows that not only is the
Galactic Plane filled with regions of high surface
brightness (i.e., regions where  Poisson noise from
background photons is elevated), but the emission is also
highly structured, having a variety of filamentary
morphologies on scales all the way down to the IRAC PSF (and
probably on smaller scales that would be significant in
data having higher angular 
resolution).  This structure further
complicates source identification and extraction.  

As a test of whether \emph{photon noise} or \emph{structure}
from diffuse backgrounds is the more significant cause of
incompleteness, we conducted the same artificial star test
described in Section 2.1 using the ``residual'' images
generated by the GLIMPSE pipeline after point sources have
been removed.  We smoothed these residual images in the
$\ell=$333\degr\ test region, first with a  101-pixel box
and then a 27-pixel box to eliminate ``structure'' on scales
smaller than about 25 times the point spread function and 7
times the PSF, respectively.  These smoothing scales
represent large and moderate smoothing, respectively.
To avoid edge-of-frame
effects, we smoothed a mosaic image of the residuals and
then reconstructed the original BCD images.  To the smoothed
residual BCD images we added Poisson noise, at a level
appropriate for each pixel, and readnoise, according to the
IRAC documentation: 11.8, 12.1, 9.1, and 7.1 electrons per
pixel for the IRAC bandpasses 3.6, 4.5, 5.8, and 8.0 $\mu$m,
respectively.  Artificial stars were added at exactly the
same positions and magnitudes as in Section 2.1 before
running the  GLIMPSE pipeline to extract sources and
construct a Catalog and Archive.  The input list of
artificial stars was compared to the recovered list of stars
to assess the completeness. 

\begin{figure} \plotone{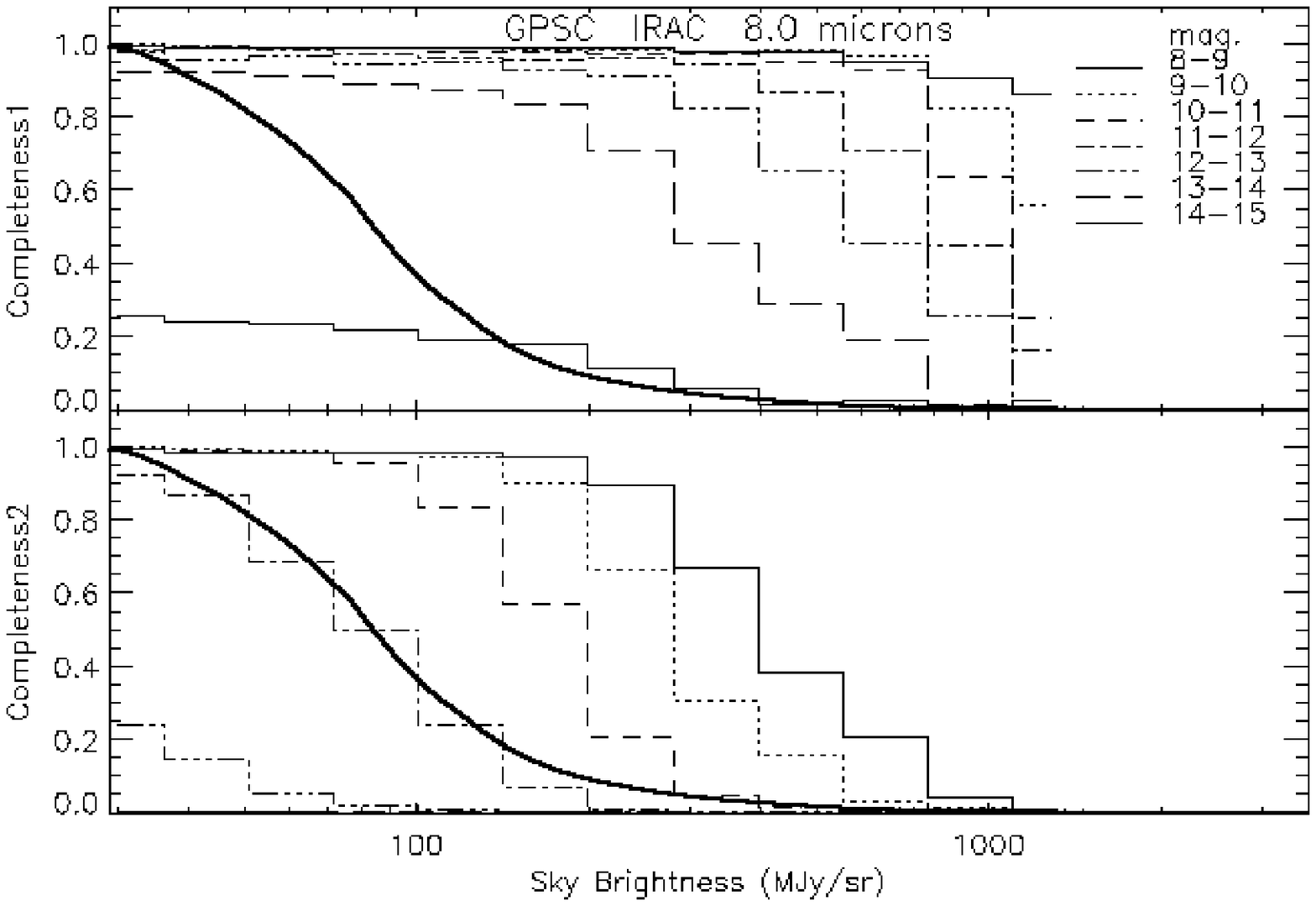} \caption{GLIMPSE I/II/3D Catalog 
Completeness1 (upper panel) and Completeness2 (lower panel)
versus sky brightness for IRAC 8.0 $\mu$m as a function of
stellar magnitude, with symbols as in Figure~\ref{comp1a}.      
\label{comp4a} } 
\end{figure}

\begin{figure} \plotone{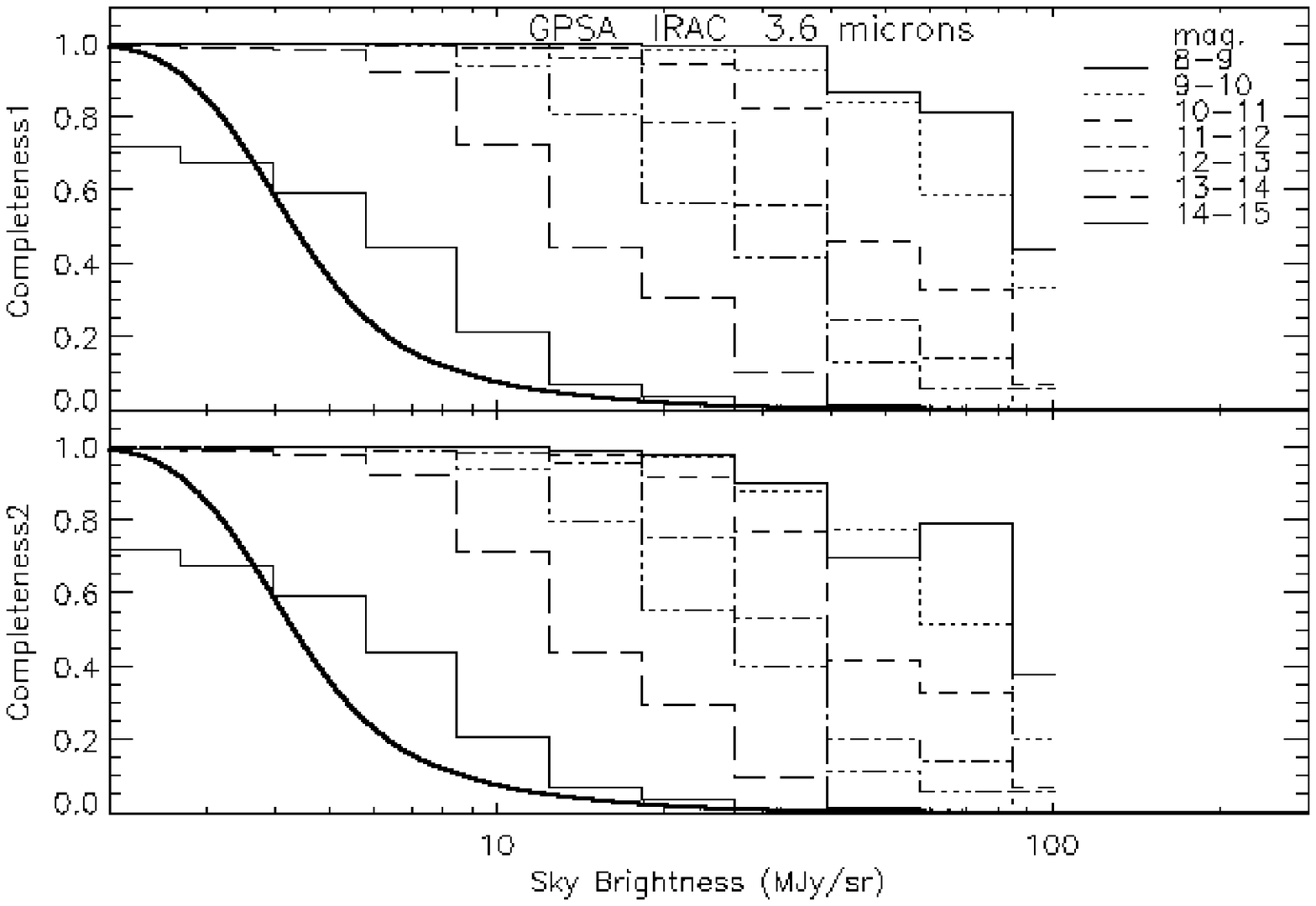} \caption{GLIMPSE I/II/3D Archive 
Completeness1 (upper panel) and Completeness2 (lower panel)
versus sky brightness for IRAC 3.6 $\mu$m as a function of
stellar magnitude, with symbols as in Figure~\ref{comp1a}.
\label{comp1aA} } 
\end{figure}

\begin{figure} \plotone{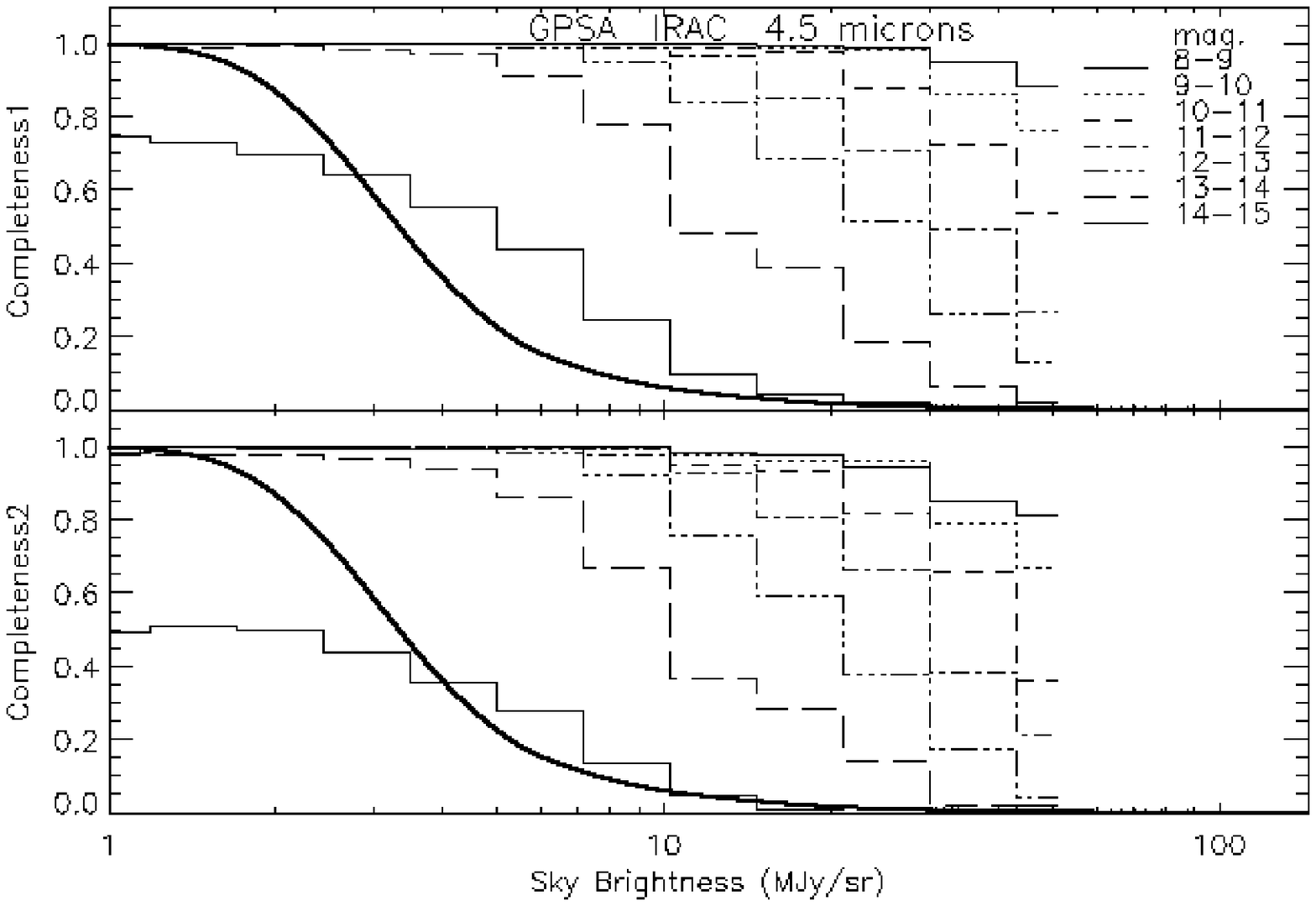} \caption{GLIMPSE I/II/3D Archive 
Completeness1 (upper panel) and Completeness2 (lower panel)
versus sky brightness for IRAC 4.5 $\mu$m as a function of
stellar magnitude, with symbols as in Figure~\ref{comp1a}.      
\label{comp2aA} } 
\end{figure}

\begin{figure} \plotone{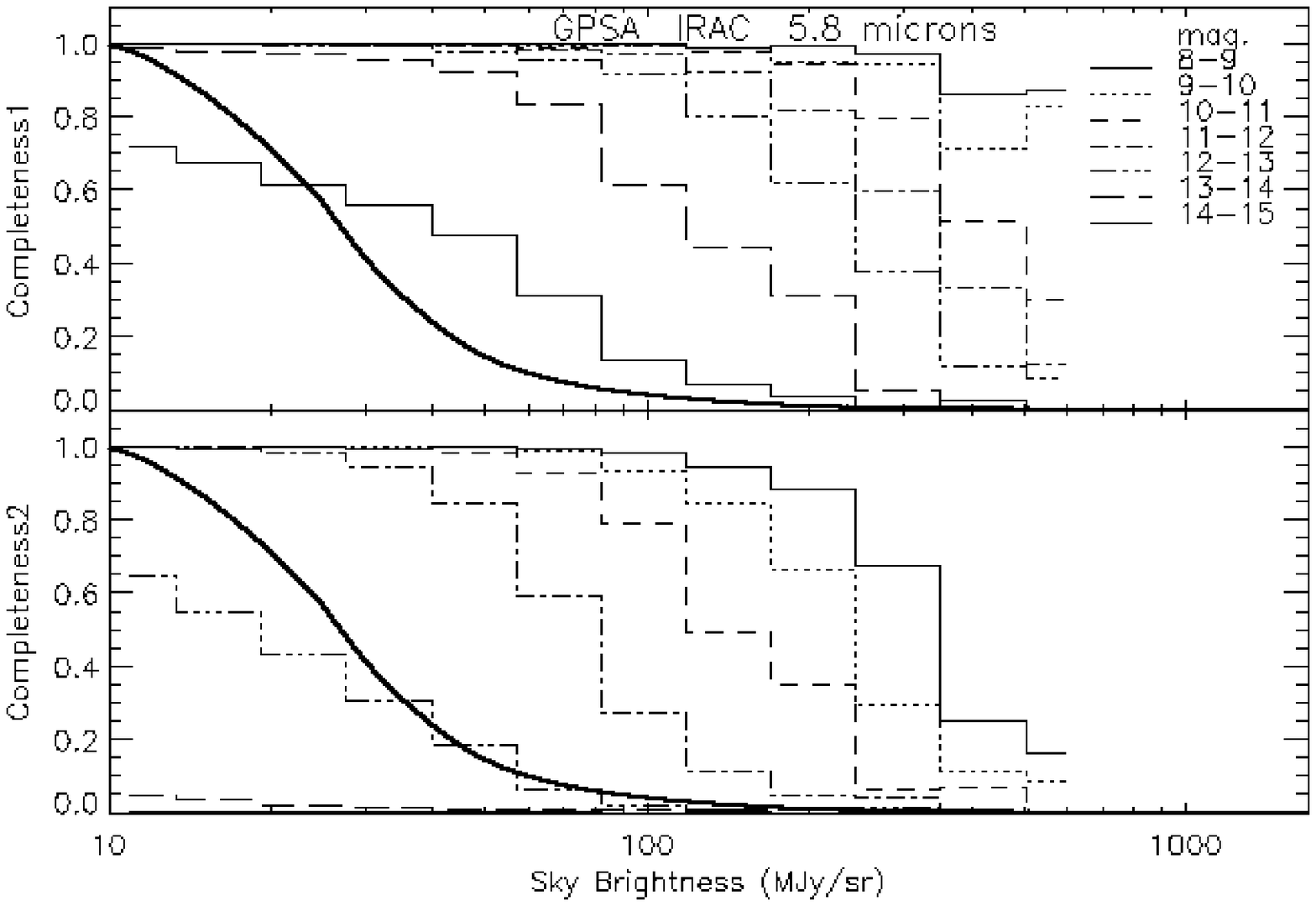} \caption{GLIMPSE I/II/3D Archive 
Completeness1 (upper panel) and Completeness2 (lower panel)
versus sky brightness for IRAC 5.8 $\mu$m as a function of
stellar magnitude, with symbols as in Figure~\ref{comp1a}.      
\label{comp3aA} } 
\end{figure}

\begin{figure} \plotone{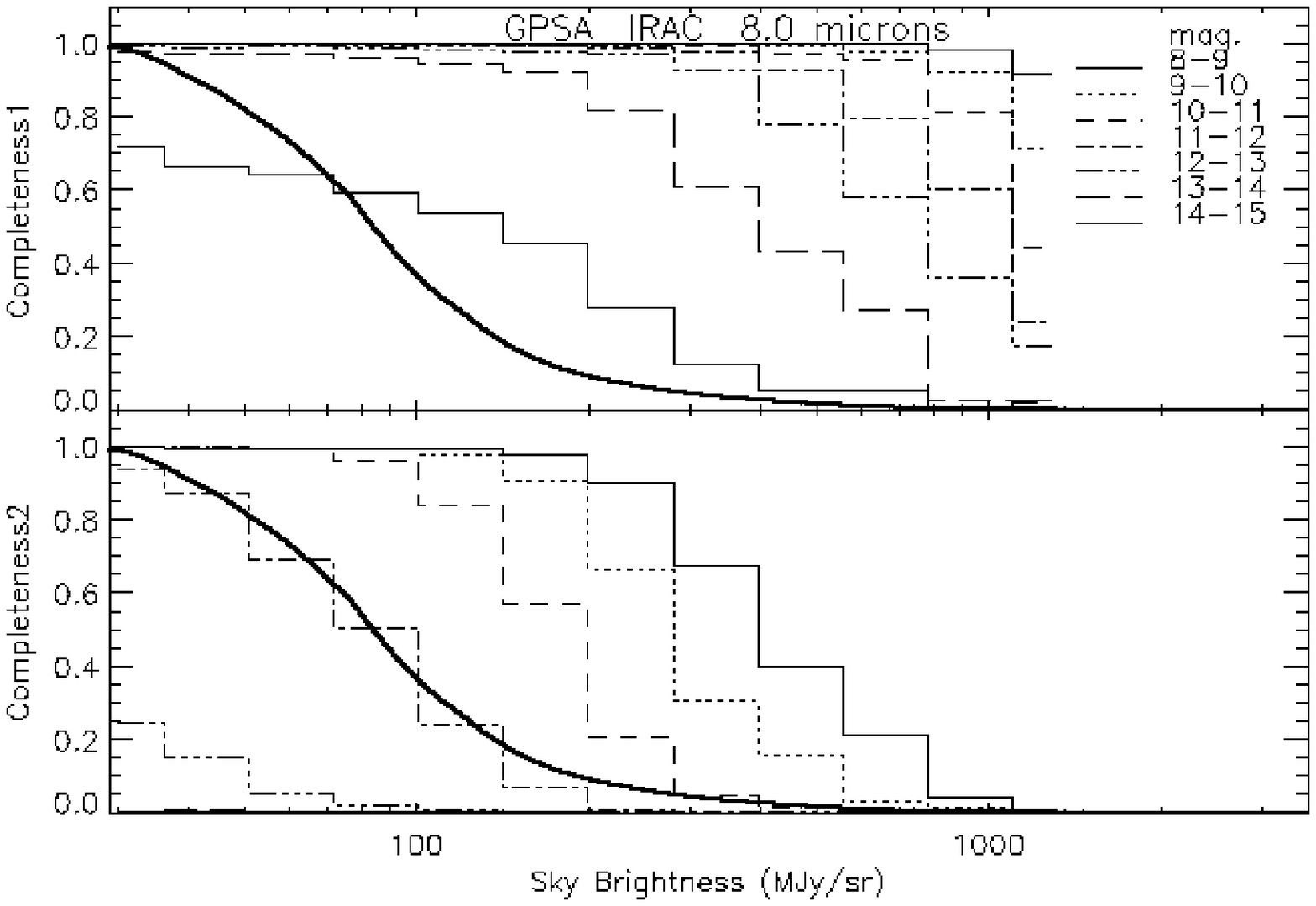} \caption{GLIMPSE I/II/3D Archive 
Completeness1 (upper panel) and Completeness2 (lower panel)
versus sky brightness for IRAC 8.0 $\mu$m as a function of
stellar magnitude, with symbols as in Figure~\ref{comp1a}.      
\label{comp4aA} } 
\end{figure}

\begin{figure} \plotone{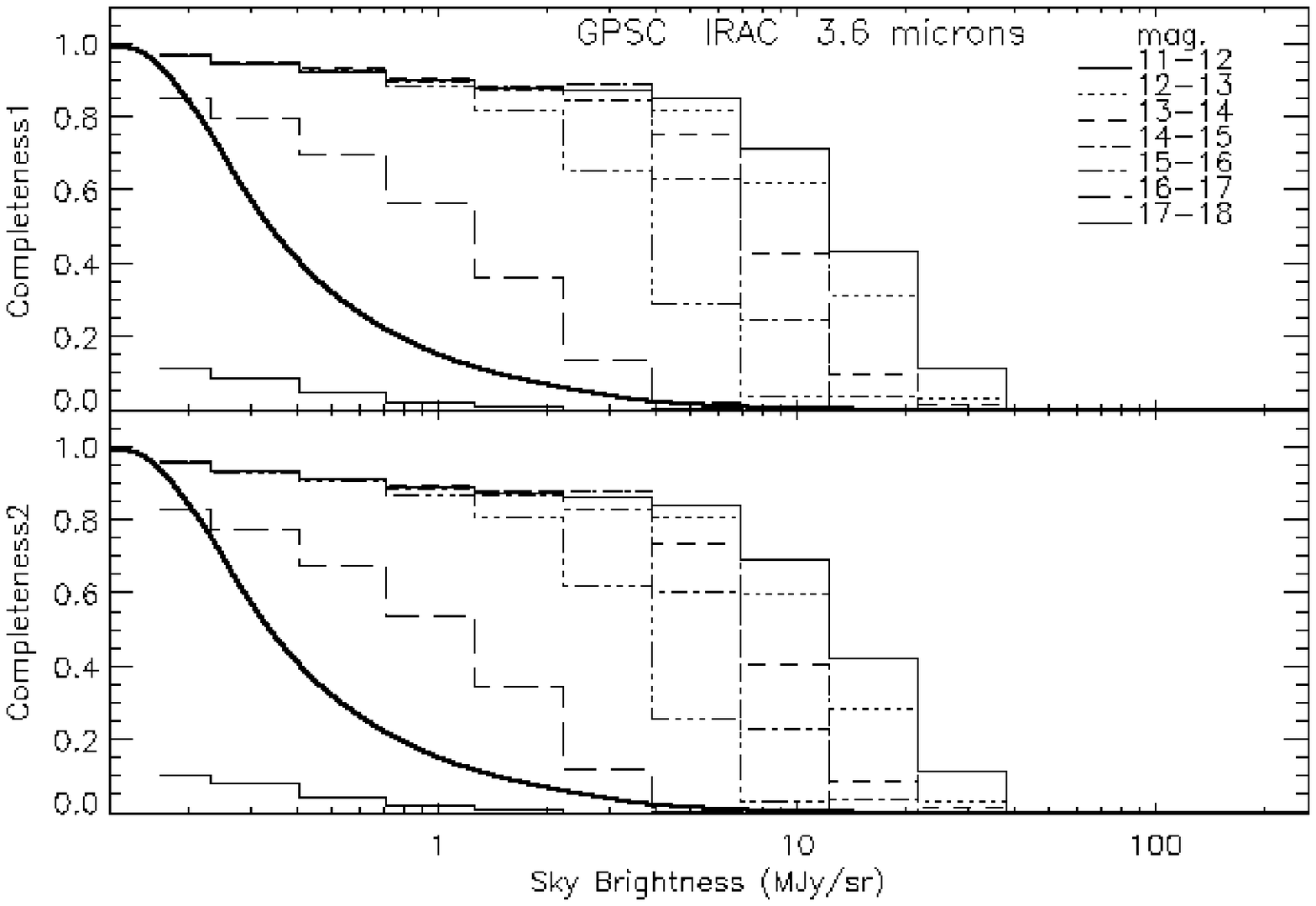} \caption{GLIMPSE 360/Deep GLIMPSE Catalog 
Completeness1 (upper panel) and Completeness2 (lower panel)
versus sky brightness for IRAC 3.6 $\mu$m as a function of
stellar magnitude, with symbols as in Figure~\ref{comp1a}.      
\label{comp5a} } 
\end{figure}

\begin{figure} \plotone{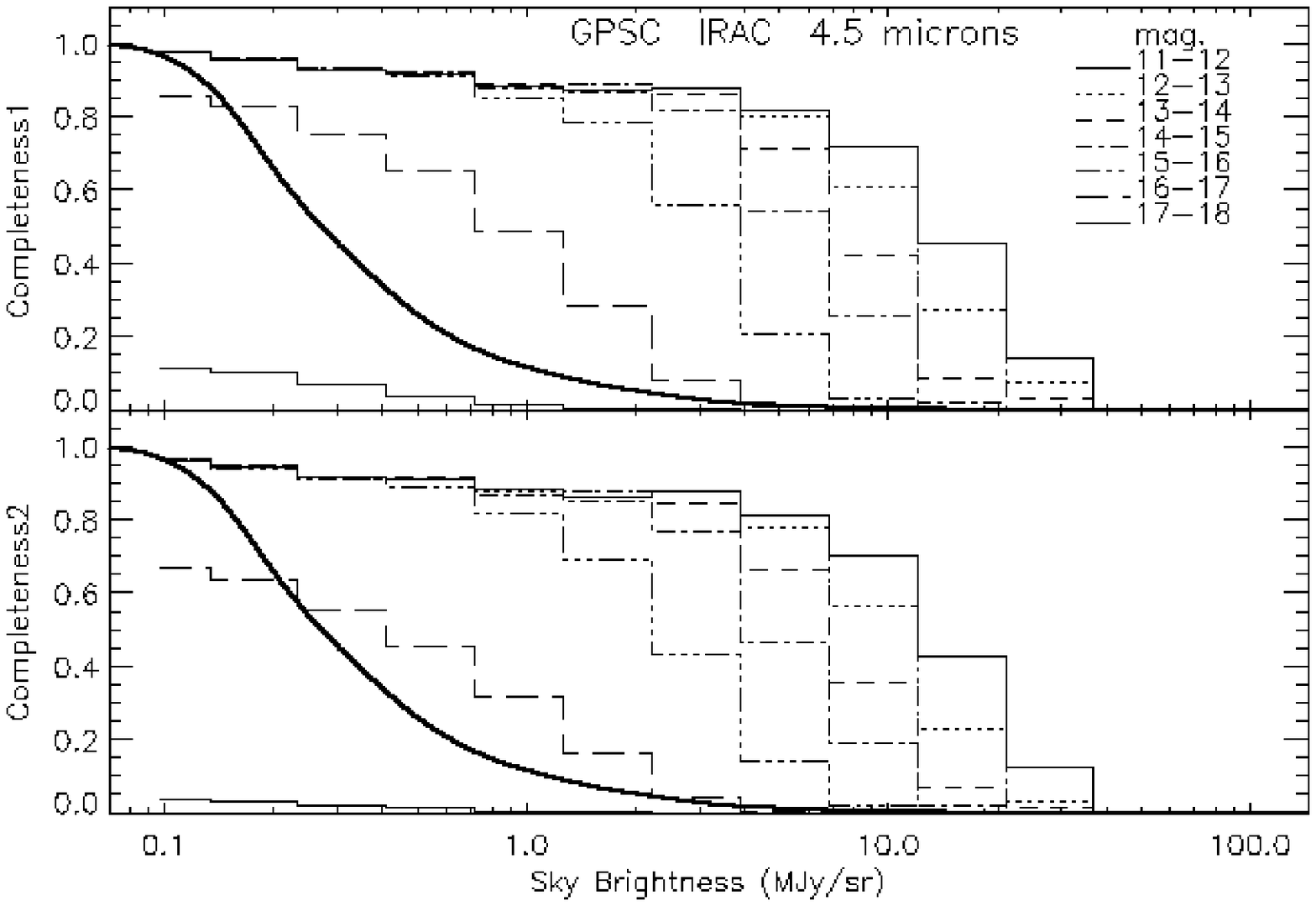} \caption{GLIMPSE 360/Deep GLIMPSE Catalog 
Completeness1 (upper panel) and Completeness2 (lower panel)
versus sky brightness for IRAC 4.5 $\mu$m  as a function of
stellar magnitude, with symbols as in Figure~\ref{comp1a}.     
\label{comp6a} } 
\end{figure}

\begin{figure} \plotone{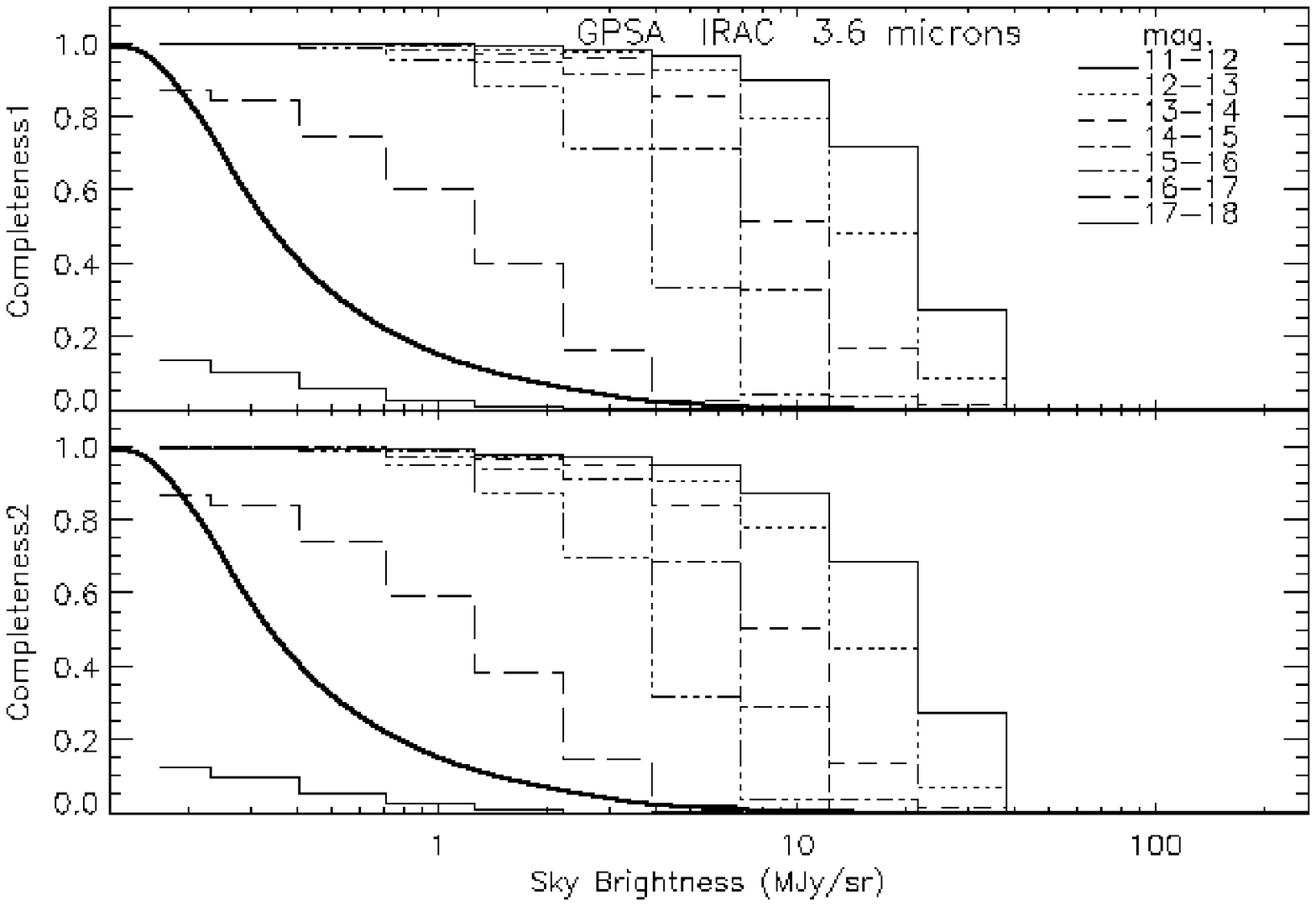} \caption{GLIMPSE 360/Deep GLIMPSE Archive 
Completeness1 (upper panel) and Completeness2 (lower panel)
versus sky brightness for IRAC 3.6 $\mu$m as a function of
stellar magnitude, with symbols as in Figure~\ref{comp1a}.      
\label{comp7a} } 
\end{figure}

\begin{figure} \plotone{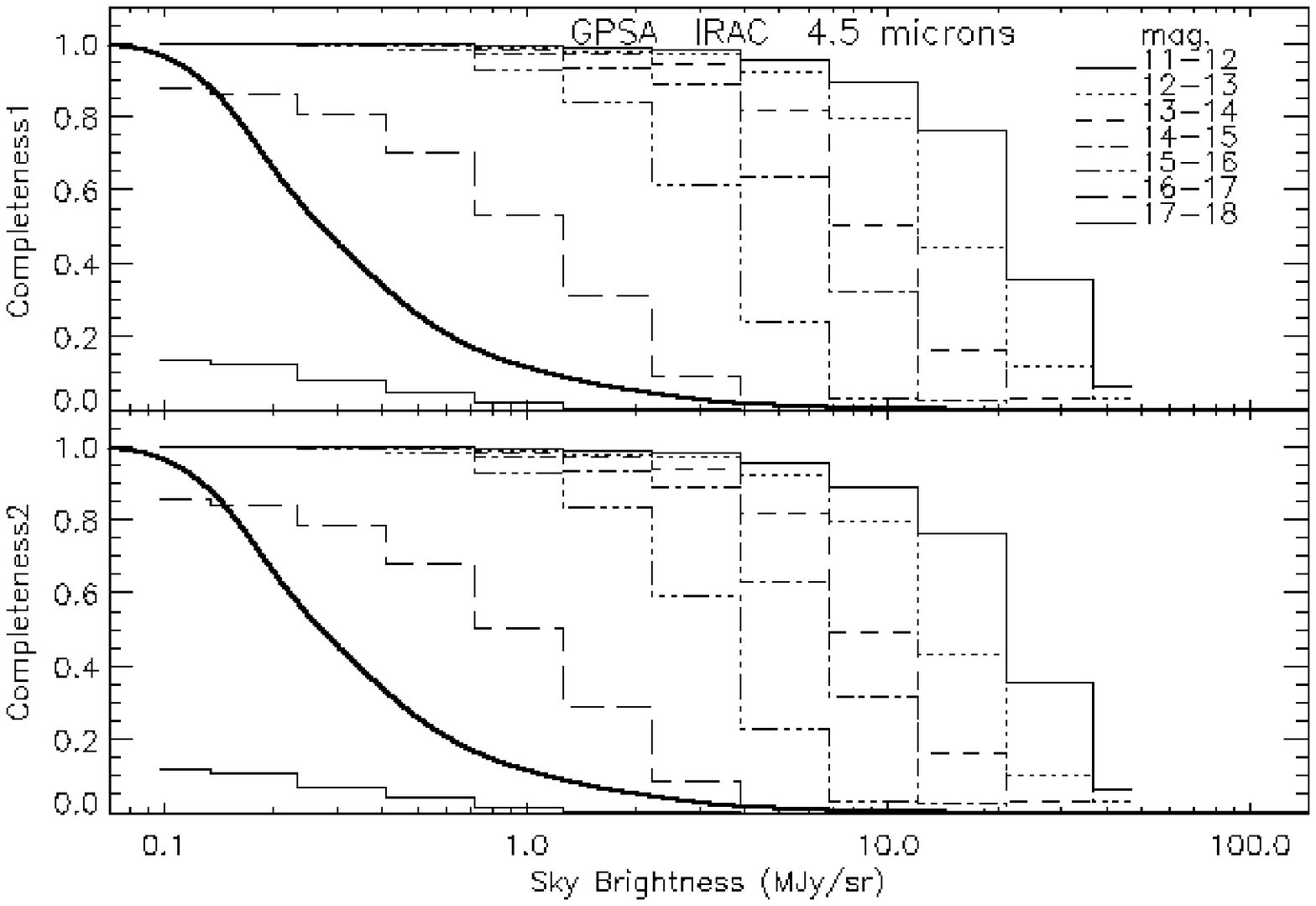} \caption{GLIMPSE 360/Deep GLIMPSE Archive 
Completeness1 (upper panel) and Completeness2 (lower panel)
versus sky brightness for IRAC 4.5 $\mu$m as a function of
stellar magnitude, with symbols as in Figure~\ref{comp1a}.     
\label{comp8a} } 
\end{figure}

\begin{figure} \plotone{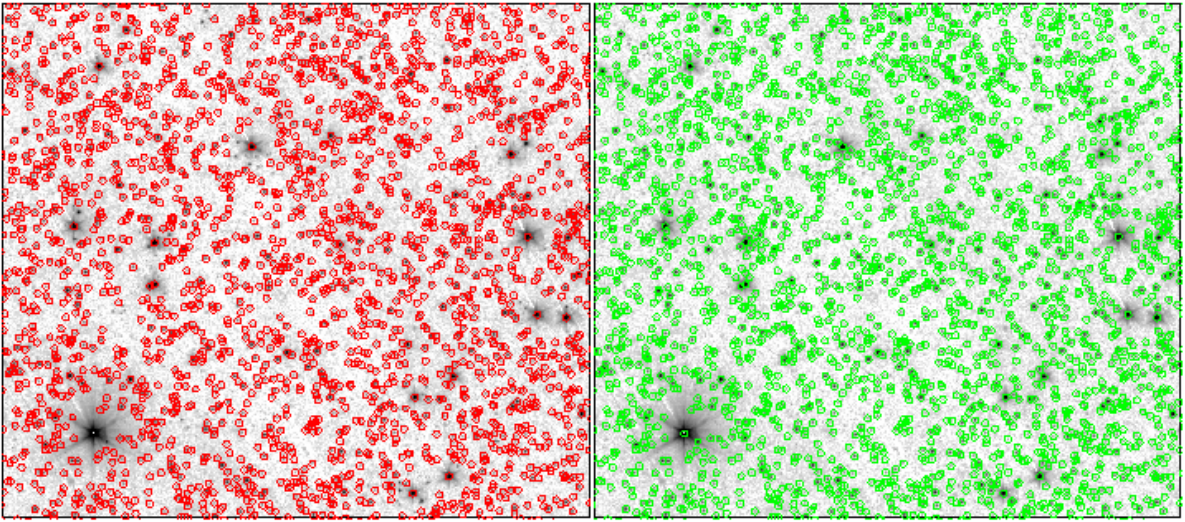} \caption{     
 Greyscale image of the GLIMPSE~360
IRAC 3.6 $\mu$m mosaic near $\ell=180$\degr. In the left panel
red circles mark the locations of GPSC sources, while in the
right panel green circles designate GPSA sources.  The subtle
differences between these figures in the vicinity of
saturated stars illustrates the greater completeness of the
Archive (right) compared to the Catalog (left).  \label{saturate180} } 
\end{figure}

\begin{figure} \plotone{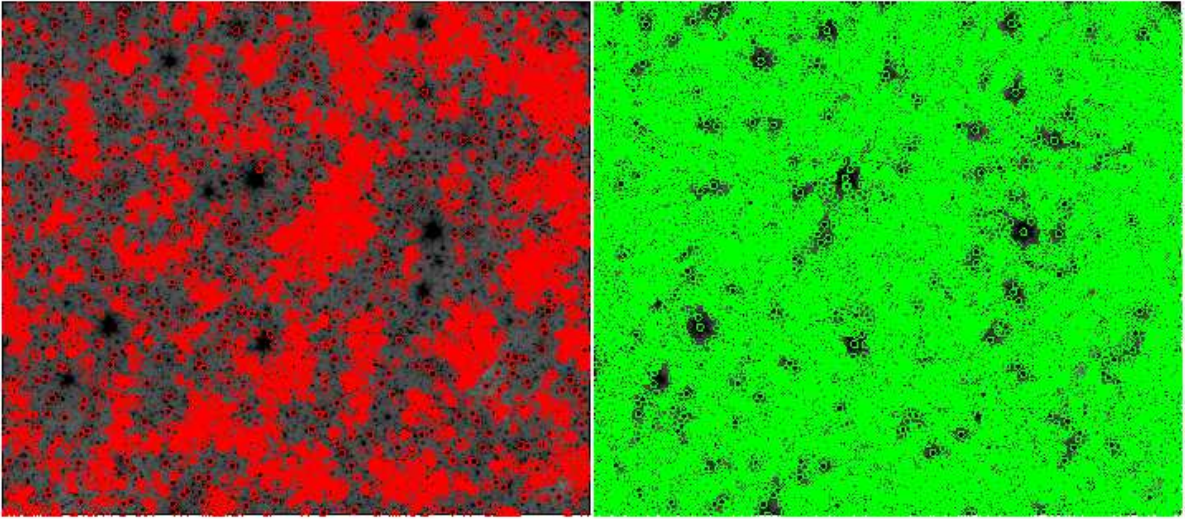} \caption{    
 As in Figure~\ref{saturate180}, but for the
crowded $\ell=30$\degr\ region of the Deep~GLIMPSE survey. 
The dramatic drop in Catalog source density (left panel)
near saturated stars illustrates the reduced completeness of
the Catalog (left) compared to the Archive (right) for such
crowded regions.  \label{saturate030} } 
\end{figure}

\begin{figure} 
\plotone{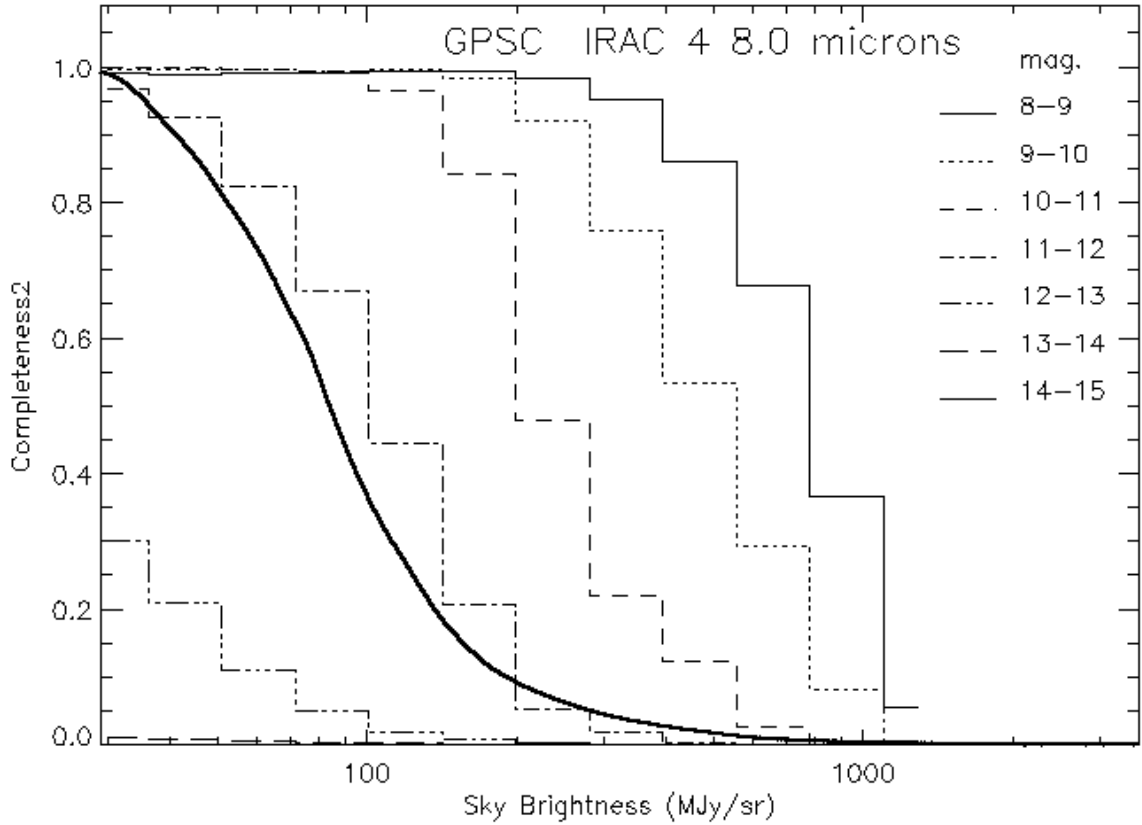} 
\caption{GLIMPSE I/II/3D Catalog Completeness2
versus sky brightness for IRAC 8.0 $\mu$m as a function of
stellar magnitude, with symbols as in Figure~\ref{comp1a}. 
The background has been smoothed using a 27-pixel boxcar.     
A comparison with Figure~\ref{comp4a} shows that
the completeness is much greater at any given magnitude range or
background level bin, indicating that background complexity
rather than photon noise is the dominant factor limiting source detection.
\label{sm27-1} } 
\end{figure}

Figure~\ref{sm27-1} shows the resulting GPSC Completeness2 versus
background level for 8.0 $\mu$m, in a way exactly analogous to
Figure~\ref{comp4a}, except that now the effect of
structure has been eliminated using a 27-pixel smoothing
box.     A comparison of Figure~\ref{sm27-1} with
Figure~\ref{comp4a} shows that the completeness is much
greater for any given magnitude or sky brightness  level
once the diffuse background has been smoothed.  Results for
other bands are similar.  

\begin{figure} 
\plotone{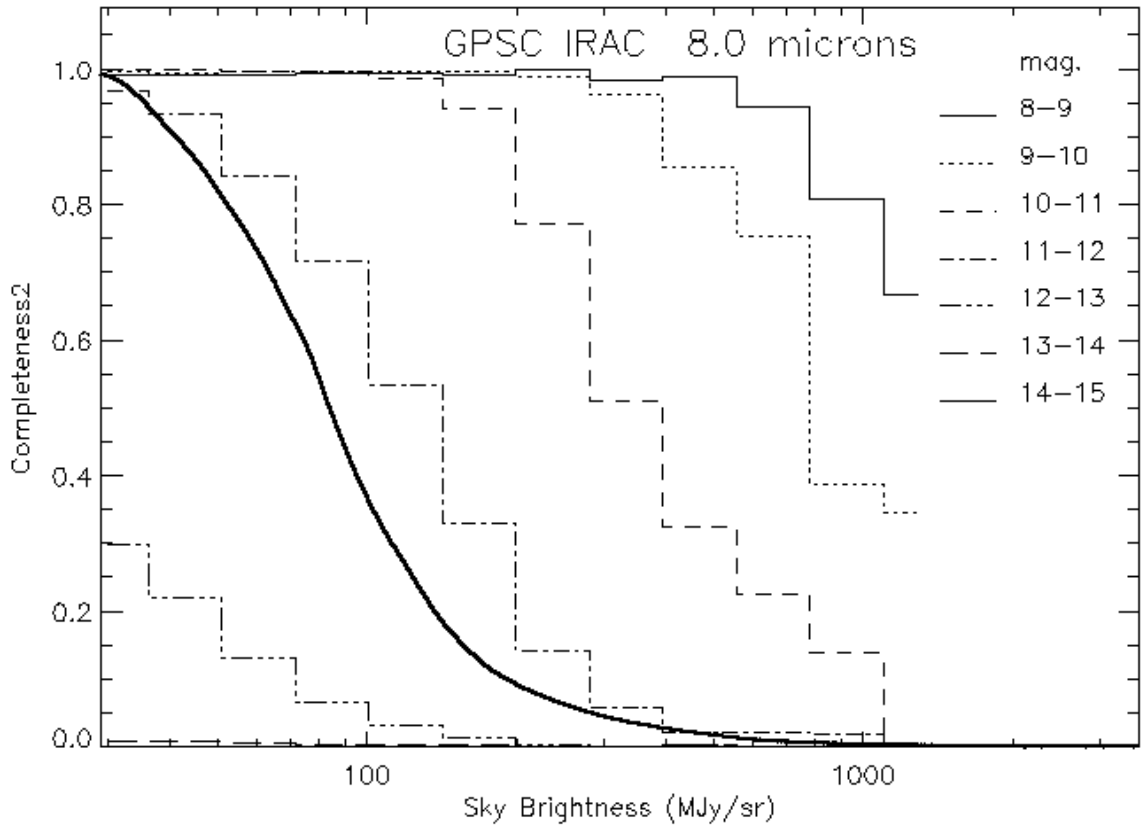} 
\caption{GLIMPSE I/II/3D Catalog  Completeness2
versus sky brightness for IRAC 8.0 $\mu$m as a function of
stellar magnitude, with symbols as in Figure~\ref{comp1a}.
The background has been smoothed using a 101-pixel boxcar.     
A comparison with Figure~\ref{comp4a} and with Figure~\ref{sm27-1} shows that
the completeness is much greater at any given magnitude range or
background level bin when the background is made yet more smooth.
\label{sm101-1} } 
\end{figure}

Figure~\ref{sm101-1} shows GPSC  Completeness2 versus
background level for the 8.0 $\mu$m band, in a way exactly analogous to
Figure~\ref{comp4a}, except that now the effect of structure
have been eliminated using a 101-pixel smoothing box. 
Comparison with Figures~\ref{comp4a} and \ref{sm27-1}
reveals that a much greater fraction of sources at all
magnitudes and background levels are recovered and recorded
in the Catalog when the diffuse background structure is
reduced even further. 

We conclude that, at the magnitude ranges of interest to
users of GLIMPSE and similar $Spitzer$ IRAC surveys,
incompleteness resulting from diffuse backgrounds is
dominated by the effects of ``structure'' in the  background
rather than purely by photon noise.  The artificial star
analysis performed in Section 2.1 implicitly includes both
effects.  In the absence of a simple, single parameter
suitable for describing the complexity of structured
background emission in the Galactic Plane, we retain the
tables and figures, as previously presented, as a way to
assess incompleteness as a function of mean background
level.

\section{Analysis of Point Source Completeness}

Point source crowding (i.e., confusion) inhibits source
detection and affects the reliability of photometry
independently from the diffuse background effects addressed
above.  Point source densities in the GLIMPSE Archives range
from $<$10 sources arcmin$^{-2}$ at 5.8 and 8.0 $\mu$m in
some local regions of the GLIMPSE I/II/3D surveys to
over 150 sources arcmin$^{-2}$ at 3.6 and 4.5 $\mu$m in some
portions the GLIMPSE 360/Deep GLIMPSE surveys. In the
first extreme, source densities are so low that their PSFs
rarely overlap and confusion effects are negligible.   In
the latter extreme, confusion is of considerable
consequence.   Ability to detect any given source  will
depend on the proximity to a neighboring source (or sources)
and on the flux ratio between the source in question and the
neighboring source(s). As such, this issue involves one more
free parameter than the diffuse background problem addressed
above.  Accordingly, we have elected to perform a more
general and more empirical characterization of its effects
using the GLIMPSE data itself rather than an artificial star
analysis.  

\begin{figure}  
\plotone{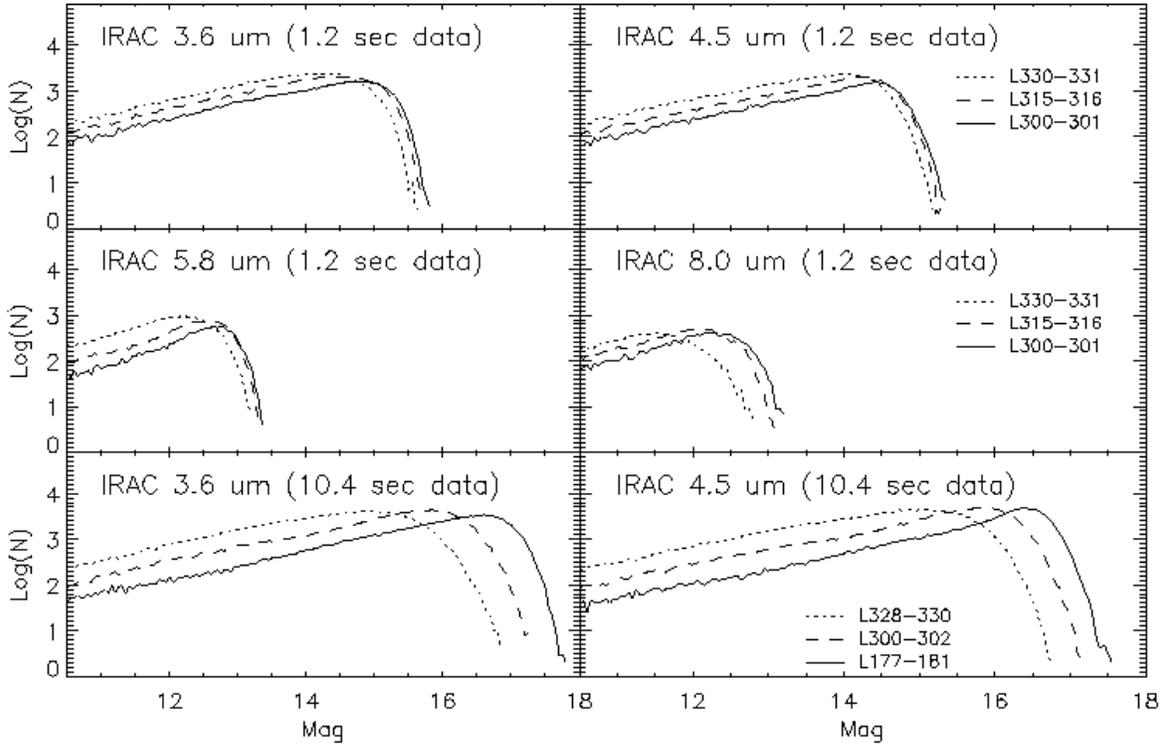}  
\caption{Source count versus magnitude plot for the IRAC GLIMPSE
I/II/3D Archive (1.2 sec exposures) and the GLIMPSE 360/Deep
GLIMPSE Archive (10.4 sec exposures).  Labels show the IRAC
bandpass in each panel, and the Y axis indicates the
log of number counts within 0.05 mag wide bins.   Line styles
demarcate distinct regions of Galactic longitude.   The
turnover in the curves indicates the magnitude at which
incompleteness  becomes significant, and the turnover point
increases (becomes fainter) with distance from Galactic
center as a result of decreasing stellar surface
densities and decreasing diffuse background levels..   
\label{lognlogs} }  
\end{figure}

Figure~\ref{lognlogs} plots the base 10 logarithm of the 
point source number counts (N) in 0.05 mag wide bins versus
magnitude (i.e., flux) from the GLIMPSE Archives.  The
$\log$(N) -- mag plot in each panel shows a different IRAC
band either for the GLIMPSE I/II/3D 1.2 s survey (all four
bands) or the GLIMPSE 360/Deep GLIMPSE 10.4~s  survey (two
bands).  Line styles denote GLIMPSE Archive data from three
distinct Galactic longitude strips, as labeled.  Only the
sources having local point source densities  in the lowest
quartile of all sources in each longitude segment were used
in order to mitigate possible effects of confusion and
ensure that results are comparable to the simulations
detailed in Section 3 where artificial stars were added at
locations devoid of known sources. The curves exhibit a
quasi-linear rise toward fainter magnitudes and then a
well-defined turnover.  The bin having the maximum counts is
designated as the ``turnover''  or ``peak'' magnitude bin
and marks where the Archive completeness becomes
appreciable, as quantified below. 
The turnover magnitude varies systematically
with Galactic longitude in the sense that regions farther
from the Galactic Center are complete to fainter magnitudes
as the source density drops.  The lower left panel shows
that the turnover magnitude toward the Galactic anticenter
is [3.6]=16.6, while at $\ell=$328--330\degr\ it is
[3.6]$\simeq$14.9. It is also clear that the  10.4 s
GLIMPSE~360/Deep GLIMPSE data go deeper than the GLIMPSE
I/II/3D surveys.  The shapes of the curves, specifically the
slopes as a function of magnitude, exhibit subtle but real
differences  reflecting variations in stellar populations
along different sightlines.  A detailed analysis of the
$\log$(N) -- mag plots can be used to infer features such as
the Galactic ``long bar''  interior to $\ell=30$\degr\ and
spiral arm tangencies 
\citep[e.g.,][]{benjamin2005,benjamin2009},  and
additional analyses will be presented elsewhere.
Nevertheless, Figure~\ref{lognlogs} serves to illustrate the
point that the completeness of the GLIMPSE surveys  as a
function of stellar surface density may be estimated from
data itself using a series of similar plots where the 
$\log$(N) -- mag curves  are generated using subsets of
Archive data selected by local source density\footnote{Local
source density (attribute SRCDENS in the GPSC and GPSA)  is
computed within the GLIMPSE pipeline for each Catalog or
Archive source, based on detections from the original BCD
frames, averaged over  1.6 arcmin scales.  The Appendix
further describes the GLIMPSE pipeline source detection
algorithm. } instead of longitude.  

\begin{figure}  
\epsscale{0.8}
\plotone{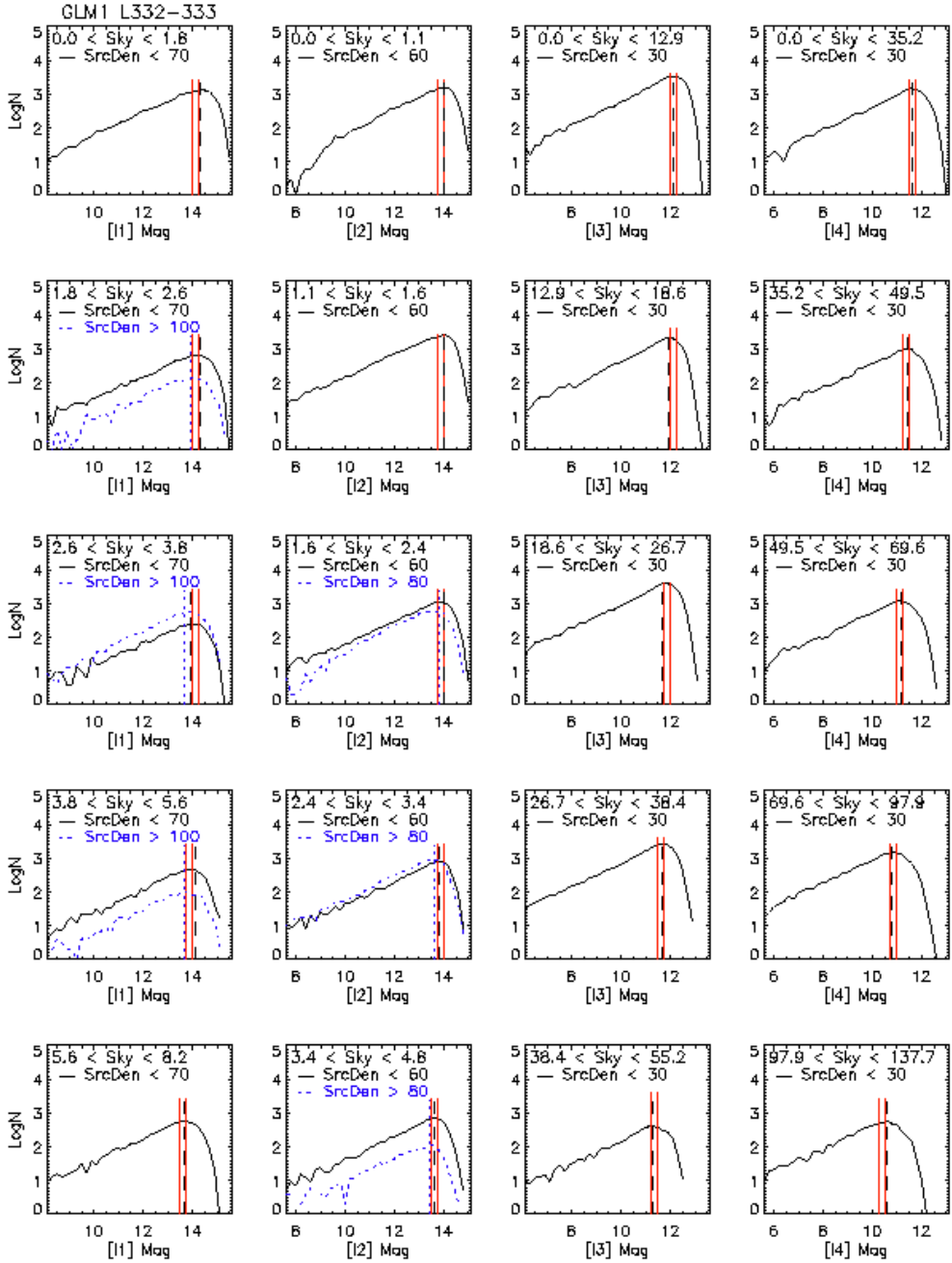}  
\caption{A grid of  $\log$(N) --
mag curves at $\ell=$332--333\degr\ with columns
showing the four IRAC bandpasses and rows
showing regions of increasing diffuse background
intensity, as labeled in each panel in units 
MJy sr$^{-1}$.  Blue dotted curves show 
source counts for regions of high
source density exceeding the labeled value in
sources arcmin$^{-2}$.  Black vertical dashed lines  
and vertical blue dotted lines mark the peak
of the $\log$(N) --
mag histograms, respectively.  Red vertical lines
designate the approximate 90\% completeness inferred
from the artificial star analysis described in Section 3.  
\label{lognlogstest1} }  
\end{figure}

Figure~\ref{lognlogstest1} shows a grid of  $\log$(N) --
mag curves from the GLIMPSE~I survey at $\ell=$332--333\degr. The four columns
correspond to the four IRAC bandpasses. The five rows
correspond sources having the specified range of diffuse background
intensity within the $\ell=$332--333\degr\ segment, 
increasing from very low values (top row) to
moderate--high values for each band (bottom row), as labeled
in each panel in MJy sr$^{-1}$.     The black solid curve
shows source counts in 0.1 mag bins for areas having local
source densities less than the indicated value, in units of
source arcmin$^{-2}$.  These correspond to relatively low 
surface densities where confusion should be minimized.   
Blue dotted curves, available in some panels, plot the
source counts in regions of high point source density.  The
vertical black dashed and blue dotted lines  mark the peak
bin for the low-density and high-density curves,
respectively.  There is a clear progression of the peaks
from right to left (from fainter  to brighter magnitudes) as
the mean sky level increases from top rows to the bottom
rows.  The blue dotted line always lies to the left
(brighter magnitudes) of the solid line, showing that
incompleteness affects brighter sources in regions of higher
source density.   Vertical red lines mark the magnitude
range of approximate 90\% completeness inferred from the
simulations reported in Section 3.  
Figure~\ref{lognlogstest1} illustrates the excellent
agreement between the vertical red lines and the vertical
black lines,  demonstrating that the peaks of the  $\log$(N)
-- mag curves are reliable indicators of the 90\%
completeness level.  As a secondary method of inferring
completeness  we fit lines to the linear part of the
$\log$(N) -- mag curve, extrapolated the line to the
location of the peak, and then measured the ratio of the
extrapolated N to the actual peak N in each plot as an
estimate of the completeness.   This method generally
supports the claim that the completeness  at the peak of
$\log$(N) -- mag curve is 70--95\%, but the
inherently non-linear nature of the curves, especially in
bright magnitude bins having few sources, precludes more
precise measurements.  

\begin{figure}  
\plotone{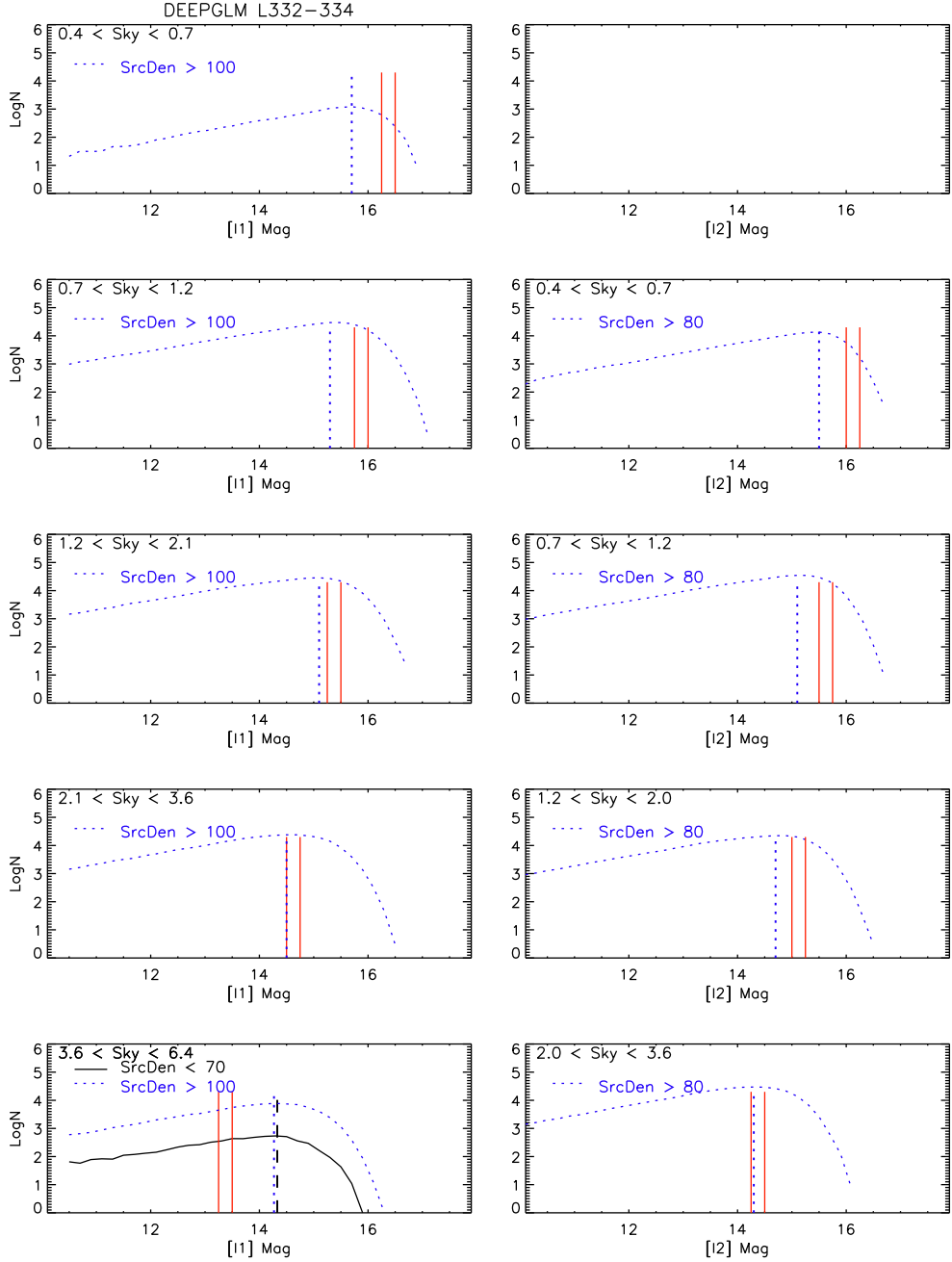}  
\caption{A grid of  $\log$(N) --
mag curves for the Deep GLIMPSE
coverage of the $\ell=$333\degr\ region,
with notation as in Figure~\ref{lognlogstest1}.
\label{lognlogstest2} }  
\end{figure}

Figure~\ref{lognlogstest2} is similar to
Figure~\ref{lognlogstest1}, but plots the $\log$(N) --
mag curves for 3.6 $\mu$m and 4.5 $\mu$m  from the
Deep GLIMPSE survey coverage of the $\ell=$333\degr\
region.   Point source densities are much higher in these
deeper 10.4 s exposures, so the dashed blue curves indicate
that the peak is shifted to brighter sources, by about
0.5--1.0 mag relative to the Deep GLIMPSE simulations (red
vertical lines) which are appropriate to uncrowded regions. 
These results indicate that the Deep GLIMPSE survey of the
inner Galaxy is limited by confusion at most locations, as expected.  As
in Figure~\ref{lognlogstest1}, the survey completeness
begins to drop at brighter magnitudes  as sky levels
increase from top to bottom.  The agreement between the blue
dotted  vertical line and the red vertical lines improves in
the lower panels.  This indicates that  diffuse background
effects become increasingly important relative to  confusion
as the sky brightness exceeds 1--2 MJy sr$^{-1}$ at both 3.6
$\mu$m and 4.5 $\mu$m.  

\begin{figure}  
\plotone{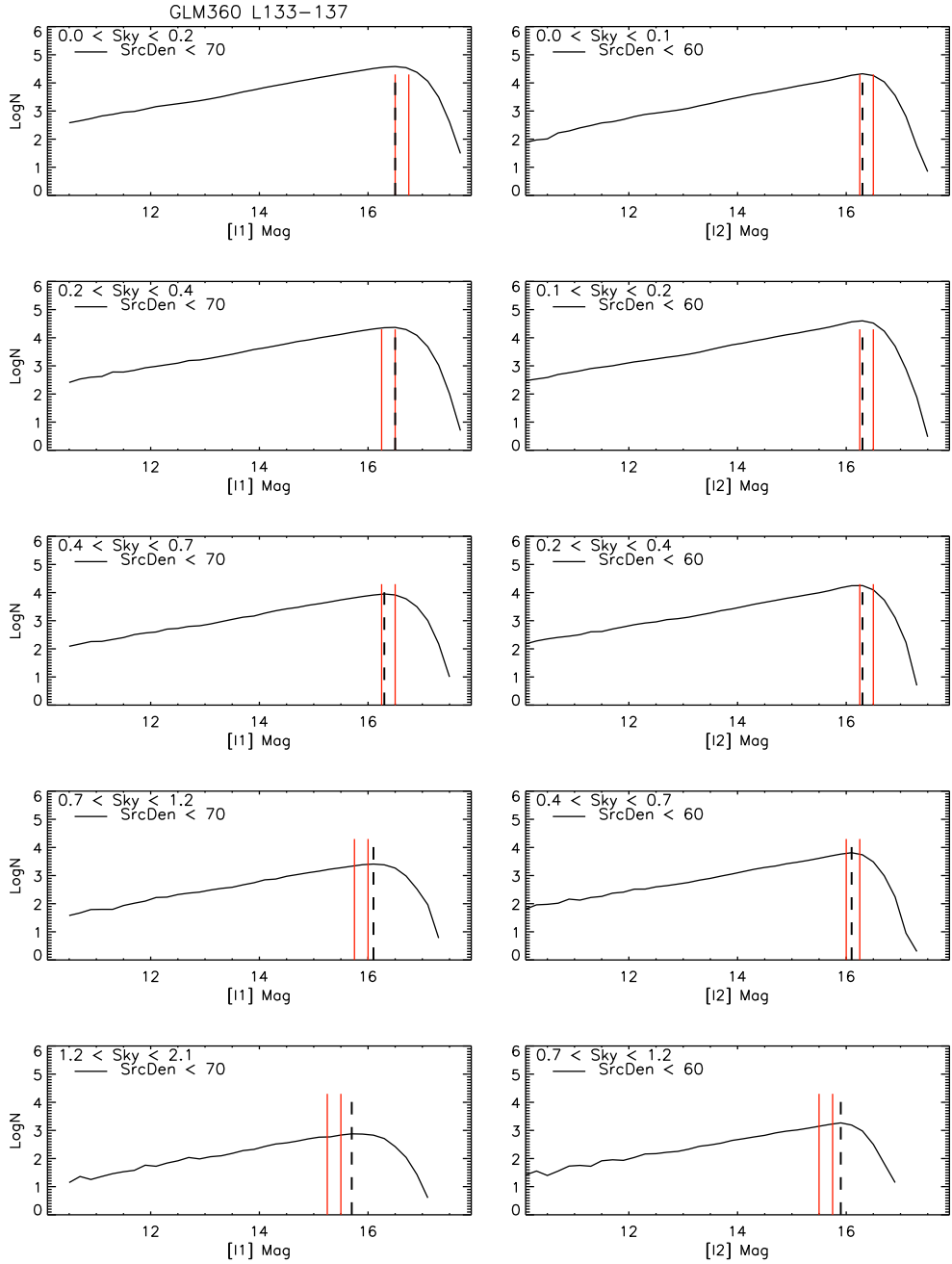}  
\caption{A grid of  $\log$(N) --
mag curves for the GLIMPSE~360
coverage of the outer-Galaxy $\ell=$133\degr\ region,
with notation as in Figure~\ref{lognlogstest1}.
\label{lognlogstest3} }  
\end{figure}

Figure~\ref{lognlogstest3} shows  $\log$(N) -- mag
curves for 3.6 $\mu$m and 4.5 $\mu$m  from the GLIMPSE~360
survey coverage of the $\ell=$133\degr\ region.  This 
outer-Galaxy sightline exhibits lower source densities and suffers
less from the effects of confusion, despite the same 10.4 s
exposure time as in Figure~\ref{lognlogstest2}.
As in Figure~\ref{lognlogstest1}, the peak
magnitude shifts toward brighter sources as
sky levels increase from top to bottom.   
There is good agreement between the completeness levels 
indicated by the simulations for GLIMPSE~360 90\%
completeness magnitudes (red vertical lines) and
the peak of the histogram (vertical dashed line), except
where sky brightness becomes large (lower rows).   
This supports the use of the peak in the $\log$(N) -- mag
as an indicator of the 90\% completeness level with regard to effects
of confusion.  
 
\begin{figure}  
\plotone{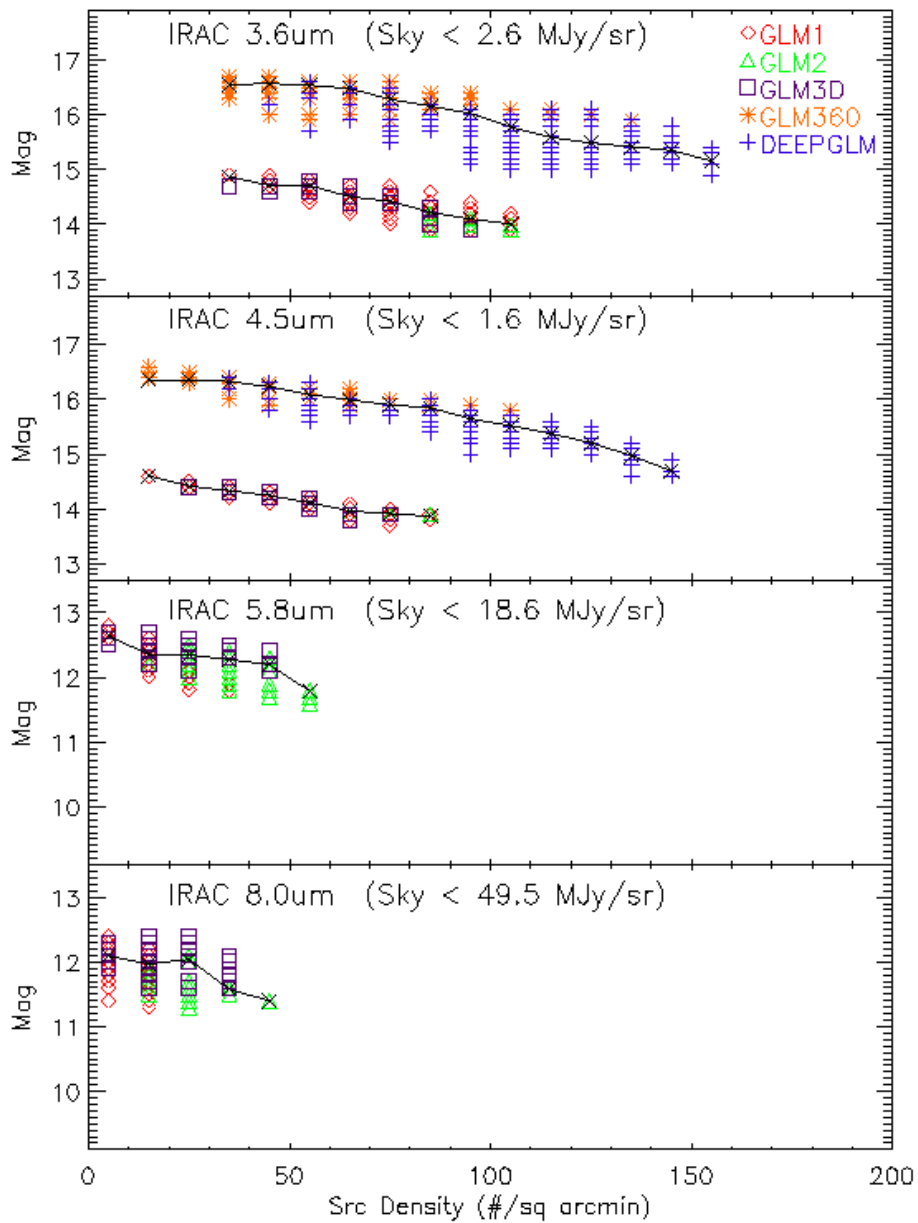}  
\caption{Limiting point
source magnitude as a function of local point source density
for each of the four IRAC bandpasses.  Source densities are
computed only for locations where the sky brightness is less
than the specified value, approximately the lowest quartile
of sky values within each three-degree GLIMPSE segment along
the Plane. Symbols differentiate the five GLIMPSE surveys. 
Lines connect the weighted means of each source density
bin.    
\label{turnover} }  
\end{figure}

Agreement between the simulated completeness levels and the
peak of the $\log$(N) -- mag histograms in the
preceding figures allows us to empirically infer the
approximate magnitude where the GLIMPSE Archives are 90\%
complete with respect to point source confusion.  
Figure~\ref{turnover} plots the $\log$(N) -- mag
turnover magnitude versus local point source density for
each of the four IRAC bands and each of the two major survey
modes.    Symbols distinguish data from each of the five
GLIMPSE legacy surveys.  At each band, only regions having
sky brightness  less than the indicated values
(corresponding to the lowest quartile of sky values in each
band) are used in order to minimize the impact of diffuse
background on point source detection.   Each data point in
Figure~\ref{turnover} represents the peak of the $\log$(N)
-- mag plot for a specified range of source densities
from 10--20 arcmin$^{-2}$ to 190--200 arcmin$^{-2}$ within a
particular segment of the GLIMPSE surveys.  In
some cases, the full range of source densities is not
realized in a given band for a given survey.  For example,
source densities never exceed 50 arcmin$^{-2}$ at 8.0 $\mu$m
(lower panel).   The dispersion of measurements at a given
source density bin reflects both the uncertainties in
measuring the turnover magnitude and real variations that
stem from the uniqueness of stellar population and
extinction distributions along each sightline. The solid
black curve connects the weighted mean turnover magnitudes
in each source density bin.    

Figure~\ref{turnover} illustrates that the limiting
magnitude is a decreasing function of source
density, most convincingly at 3.6 and 4.5 $\mu$m where
stellar photospheres are brightest  and source densities are
highest.   On average, the GLIMPSE 360/Deep GLIMPSE surveys
go almost two magnitudes deeper than the GLIMPSE I/II/3D
surveys, but the slopes are similar.   The dispersion within
a given bin is notably large in the Deep GLIMPSE  data at
3.6 $\mu$m. In the lower two panels source densities almost
never exceed 50 arcmin$^{-2}$, and the trends are less
clear.   The limiting magnitudes are nearly constant at
[3.6]$\simeq$12.4 and [4.5]$\simeq$12.1 until the  two highest
density bins where the statistics become poor.  The
dispersion at a given source density is especially large for
the 8.0 $\mu$m  panel, reflecting effects from variable
levels of diffuse sky emission that cannot be completely
removed by limiting the analysis to the lowest quartile of
sky backgrounds.   Figure~\ref{turnover} provides the best 
general estimate yet available for estimating point source
completeness levels in GLIMPSE data products, as affected by confusion.  The results
should be broadly applicable to regions covered by the
various GLIMPSE surveys.    Users of GLIMPSE data products
requiring high levels of completeness precision should conduct their own
analysis in specific regions of interest by constructing $\log$(N) -- mag
curves and making use of the SRCDENS attribute provided
as part of the GLIMPSE Catalogs and Archives.

\subsection{Limiting magnitude of GLIMPSE
data products: diffuse background versus point source confusion}

\begin{figure}  
\plotone{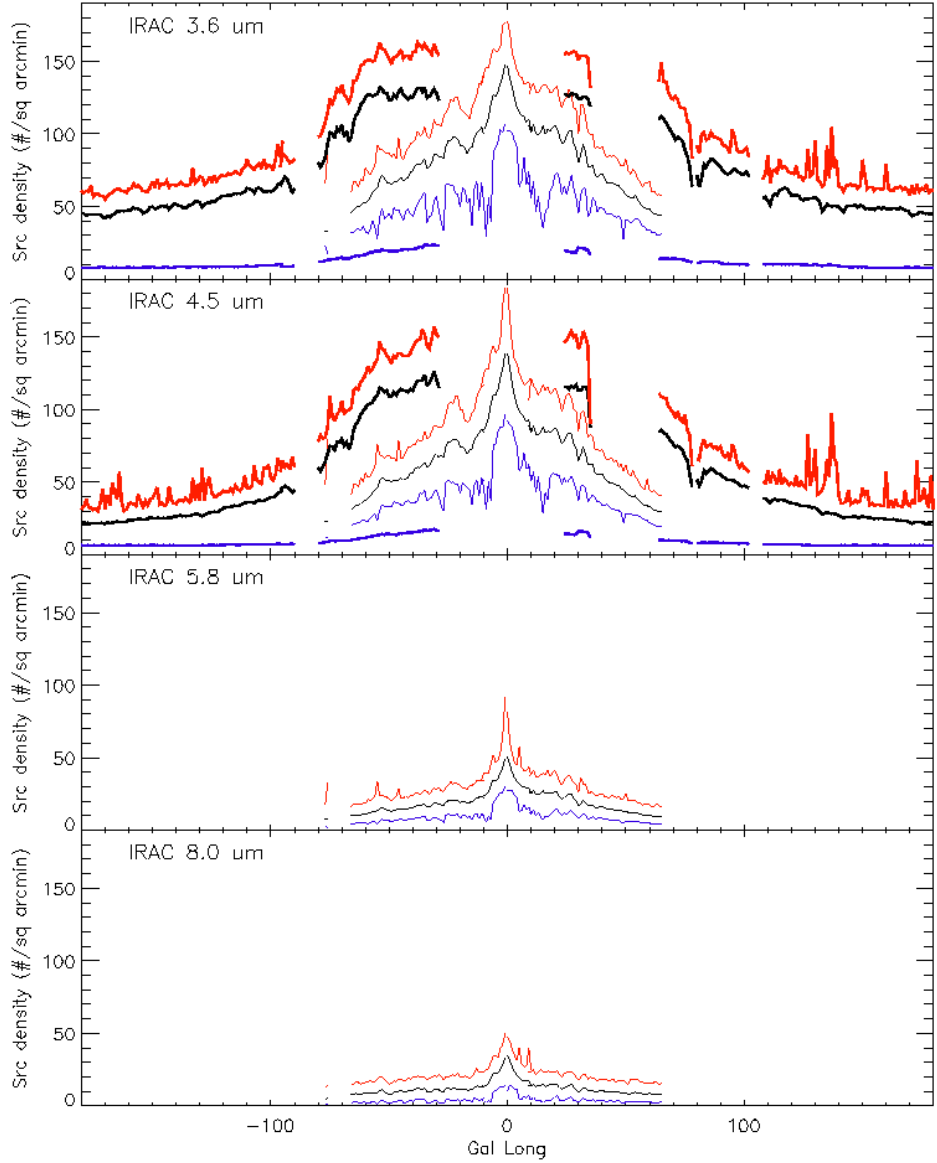}  
\caption{Point source density as a function of Galactic Longitude for
the GLIMPSE I/II surveys (thin lines) and the GLIMPSE~360/Deep GLIMPSE surveys
(bold lines).  The two sets of lines 
may not agree at overlapping longitudes because
latitude coverage of the surveys differs.   
Red/black/blue colors denote the 99.9\%/median/0.1\%
source densities, respectively.   
\label{gal_pnt} }  
\end{figure}

\begin{figure}  
\plotone{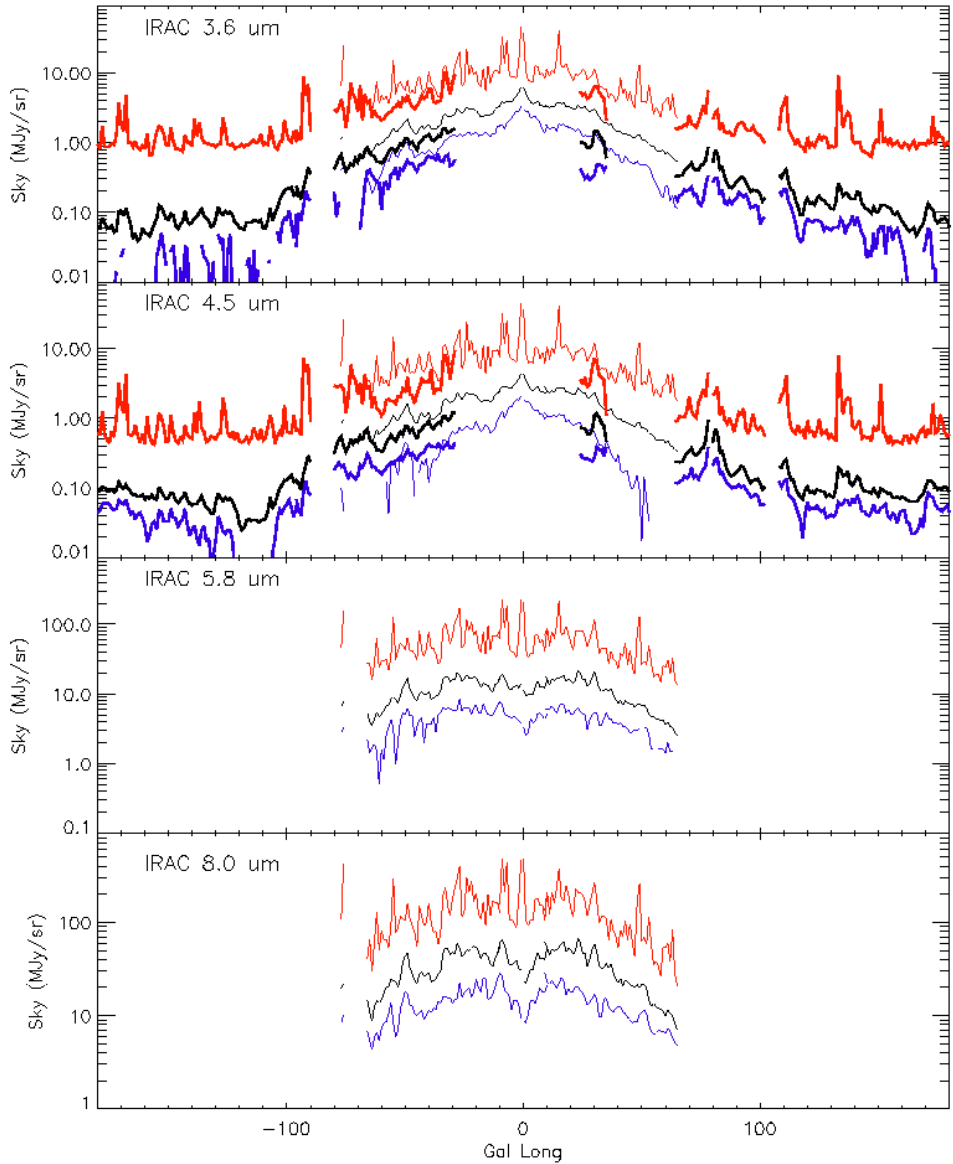}  
\caption{Diffuse background level as a function of Galactic Longitude for
the GLIMPSE I/II surveys (thin lines) and the GLIMPSE~360/Deep GLIMPSE surveys
(bold lines). The two sets of lines 
may not agree at overlapping longitudes because
latitude coverage of the surveys differs. 
Red/black/blue colors denote the 99.9\%/median/0.1\%
sky levels, respectively.
\label{gal_bkg} }  
\end{figure}

Consumers of GLIMPSE data products are advised that either 
diffuse background emission or point source confusion (or
both!)  limits the completeness of the GLIMPSE data
products  in various regimes.   Figures~\ref{gal_pnt} and
\ref{gal_bkg} plot point source densities and diffuse
background levels, respectively, as a function of Longitude
for each of the IRAC bands.  Thin lines show the GLIMPSE
I/II survey data and bold lines show the GLIMPSE~360/Deep
GLIMPSE survey data.   Red/black/blue colors denote the
99.9\%/median/0.1\% points of the distribution,
respectively, at a given longitude.  Discrepancies between
surveys at the same longitude are  the result of different
exposure times and different latitude ranges for the
GLIMPSEI/II surveys relative to  the Deep GLIMPSE survey.  

Figures~\ref{gal_pnt} and \ref{gal_bkg} show that
point source densities and sky levels generally increase
toward the Galactic center.  Median point source densities above
50 sources arcmin$^{-2}$ at 3.6 and 4.5 $\mu$m  are common
in the 1.2 s GLIMPSE I/II/3D surveys toward the inner Galaxy, 
but at 5.8 and 8.0 $\mu$m densities rarely exceed 30 sources
arcmin$^{-2}$.   Median sky brightness levels vary between
0.1 and 2 MJy sr$^{-1}$ toward the inner Galaxy at 3.6 and
4.5 $\mu$m, but at 5.8 and 8.0 $\mu$m levels between 10 and
100 MJy sr$^{-1}$ are common. The outer Galaxy, by contrast,
shows vastly reduced point source densities and diffuse sky
levels.   Source densities and diffuse levels also increase
toward the mid-Plane, but these variations are not reflected
in the latitude-averaged plots. Spatial variations in these
quantities can be large over very small angular scales.  
For example, infrared dark clouds  exhibit some of the
lowest sky levels of $<$1 MJy sr$^{-1}$, and these 
often  lie adjacent to bright-rimmed clouds and
photodissociation regions  with sky levels exceeding 100 MJy
sr$^{-1}$.  This means that a detailed description of how confusion
or diffuse background affects photometry as a function of
sky position would require a more complex analysis than
undertaken here,  but our simulations provide 
some insights specific to individual bandpasses.  

\subsubsection{GLIMPSE I/II/3D}

{\it 3.6 $\mu$m} --- Figure~\ref{turnover} shows that, for
regions with very low sky background $<$2.6 MJy sr$^{-1}$,
the GLIMPSE I/II/3D Archives are complete to between [3.6] =
14 -- 14.9 mag, depending on local source density.   By
comparison,  Figure~\ref{comp1aA} (lower panel) reveals that
completeness never exceeds 80\% in this faintest 
magnitude bin, even at low sky levels, and it
drops markedly in regions with sky brightness above 6 MJy
sr$^{-1}$.  Diffuse background significantly impacts 
completeness in brighter magnitude bins when it exceeds 10
MJyr sr$^{-1}$.  We conclude that for regions with sky
levels $\lesssim$3 MJy sr$^{-1}$, the limiting magnitude is set
by confusion but is otherwise dominated by diffuse
emission.

{\it 4.5 $\mu$m} --- Figure~\ref{turnover} shows that, for
regions with background levels $<$1.6 MJy sr$^{-1}$, the
GLIMPSE I/II/3D Archives are complete to [4.5] = 14 -- 14.7
mag, for source densities from 85 to 15 arcmin$^{-2}$.  
Figure~\ref{comp3aA} (lower panel), by comparison, shows
that the Archives are, at most, 50\% complete for the
[4.5]=14--15 mag bin even in the darkest regions.   For sky
levels exceeding about 6 MJy sr$^{-1}$, incompleteness
becomes appreciable in the  [4.5]=13--14 mag range.  We
conclude that for the darkest regions and highest source
densities completeness is limited by confusion but for  sky
levels $\gtrsim$6 MJy sr$^{-1}$, diffuse background becomes the
limiting factor.

{\it 5.8 $\mu$m} --- Figure~\ref{turnover} shows that the
limiting magnitude is roughly constant at  [5.8] = 12.3
across the range of observed source densities for sky
brightness  $<$18.6 MJy sr$^{-1}$. Figure~\ref{comp4aA}
(lower panel) indicates that  the Archives are already no
better than 55\% complete in this magnitude range, and that
by 40 MJy sr$^{-1}$, the [5.8]=11--12 mag range has become
substantially incomplete.    We conclude that above 40 MJy
sr$^{-1}$ the incompleteness is dominated by diffuse
emission at all magnitude ranges.     In the darkest areas
having $<$18.6 MJy sr$^{-1}$  the effects of confusion and
diffuse background may be comparable for sources [5.8]=
12--13.   Given the very low source densities and strong PAH emission
at 5.8 $\mu$m we conclude that diffuse emission often 
limits the detection of point sources for typical 
Galactic Plane sightlines.  

{\it 8.0 $\mu$m} --- Figure~\ref{turnover}  shows that the
limiting magnitude is approximately constant  at
[8.0]$\simeq$12 across the small range of observed source
densities for sky values $<$49.5 MJy sr$^{-1}$.   
Figure~\ref{comp2aA} (lower panel) indicates that in the
[8.0]=11--12 range, the completeness has already dropped
below 90\% for sky values near 49.5 MJy sr$^{-1}$.   By 100
MJy sr$^{-1}$ stars in the 10--11 mag range become
substantially incomplete.   Given the very low source
densities and strong PAH emission at 8.0 $\mu$m we conclude
that diffuse emission is frequently the limiting agent
regarding the detection of point sources.  

\subsubsection{GLIMPSE 360/Deep GLIMPSE}

{\it 3.6 $\mu$m} --- Figure~\ref{turnover}  displays a range
of limiting magnitudes from 16.6 at the lowest source
densities (15 arcmin$^{-2}$) to 15.2 at densities approaching
155 arcmin$^{-2}$ for sky backgrounds $<$2.6 MJy
sr$^{-1}$.   By comparison, Figure~\ref{comp7a} reveals
that  at these sky levels stars in the [3.6]=15--16 bin are
already less than 90\% complete. We conclude that the
effects of confusion are dominant for sightlines with the
highest stellar densities and lowest backgrounds, but that
diffuse emission limits completeness  for many sightlines
and magnitude ranges.  Diffuse sky background becomes a
signficant/limiting factor above  about 1--2 MJy sr$^{-1}$
at both   3.6 and 4.5 $\mu$m. 

{\it 4.5 $\mu$m} --- Figure~\ref{turnover} illustrates that
limiting magnitudes vary from [4.5]=16.3--14.7 as source
densities increase from 15 to 145 arcmin$^{-2}$  for sky
backgrounds $<$1.6 MJy sr$^{-1}$.   By comparison,
Figure~\ref{comp8a} shows that  stars in the [4.5]=15--16
mag bin are already less than 90\% complete at these sky
levels.  We conclude that sky background effects dominate
the incompleteness unless sky levels are $\lesssim$2 MJy sr$^{-1}$
and point source densities are $\gtrsim$80 sources arcmin$^{-2}$.

\section{Conclusions}

We have conducted an analysis of the completeness of point
sources appearing in the various GLIMPSE Point Source
Catalogs and Archives to assess the impact of 1) high diffuse
Galactic backgrounds and 2) high point source densities.     
With regard to the effects of diffuse backgrounds, we find:

\begin{enumerate}

\item{Completeness is, as expected, a strong function of both stellar
magnitude and background sky level, such that no simple
characterization of the completeness is possible.
Completeness drops rapidly with increasing stellar magnitude
and sky brightness, especially in the IRAC 5.8 and 8.0 $\mu$m bands
where strong PAH features produce large sky background
brightnesses.}

\item{We provide figures and tabular data, suitable for
interpolation, listing Completeness1 and Completeness2
values for each band for each of the two GLIMPSE survey
strategies for both the Catalogs and Archives. }

\item{Completeness, as defined in two slightly different
ways,  is generally higher in the GLIMPSE Point Source
Archives than in the Catalogs, at any given IRAC band. 
Completeness2 is always equal to or lower than Completeness1.}

\item{Catalogs resulting from the GLIMPSE~360/Deep GLIMPSE
surveys may suffer significantly reduced completeness
relative to the equivalent Archives because sources in the
PSF wings of saturated stars are removed from the Catalogs. 
This effect is most severe in high-density regions toward
the Inner Galaxy.  Consumers of GLIMPSE~360/Deep GLIMPSE
data products should consider using the Archive rather than
the Catalog if high completeness is desired.  GLIMPSE
I/II/3D surveys, by virtue of their shorter exposure times,
exhibit a relatively minor completeness differences between
Catalogs and Archives from this effect. }

\item{We assess whether background ``brightness'' or
background ``structure'' is the limiting factor for point
source detection.  We conclude that both play a role, but
that the effects of structure appear to dominate over pure
photon noise in hindering the identification and extraction
of point sources. }

\item{Although our artificial star recovery analysis is performed
on images drawn from the GLIMPSE~I/II/3D (multiple
1.2-second exposures at all four IRAC bands) and GLIMPSE~360
and Deep GLIMPSE (three 10.4-second exposures at 3.6
and 4.5 $\mu$m)  surveys, the results can be generalized to
estimate completeness for other {\it Spitzer} IRAC programs that
used similar observing strategies and have data products
(Catalogs and Archives) constructed in the same manner.
These include the SMOG, Cygnus-X, and the Vela-Carina surveys.}

\end{enumerate}

With regard to the effects of high stellar density on
completeness, we conclude that only in the regions of low
sky background do the effects of confusion limit the
detection of point sources in the shallow GLIMPSE I/II/3D
surveys.   The relative importance of  diffuse background
emission increases from 3.6 $\mu$m to 8.0 $\mu$m for a
typical sightline  and as the point source densities
decrease, such as toward the Outer Galaxy. In GLIMPSE
I/II/3D, confusion is the dominant source of incompleteness
only at 3.6 and 4.5 $\mu$m and in regions  of high point source
density $\gtrsim$80 sources arcmin$^{-2}$,  and where sky
background levels are $\lesssim$2 MJy sr$^{-1}$.  IRAC 5.8
and 8.0 $\mu$m are almost never limited by confusion as
point source densities never exceed about 50 sources
arcmin$^{-2}$. GLIMPSE~360/Deep GLIMPSE surveys reach
several magnitudes deeper and are limited by both confusion
and diffuse emission, to varying degrees.  Confusion is the
dominant source of incompleteness  at 3.6 and 4.5 $\mu$m and
regions  of high point source density $\gtrsim$80 sources
arcmin$^{-2}$,  and where sky background levels are
$\lesssim$2 MJy sr$^{-1}$.  Figure~\ref{turnover} provides
the best current estimates for confusion limits in each of
the GLIMPSE surveys and bandpasses as a function of point
source surface density.   GLIMPSE data products include
attributes for quantifying local point source density and
diffuse sky levels for each detected source.  

Finally, we caution that completeness in the GLIMPSE data
products (and any higher-level data catalog products)  is
inevitably also affected  by additional selection and
matching criteria that are imposed to ensure high
reliability (i.e., robustness against false sources).  
Therefore, completeness is necessarily a  multi-faceted,
non-trivial quantity that is unique to a particular survey
strategy and source extraction paradigm.   Consumers of
GLIMPSE data requiring highly precise completeness estimates are urged to 
understand the details of the source detection and selection
process for the data products in question.

\acknowledgments 

We are grateful to Matt Povich,  Bob Benjamin,  and Dan
Clemens for their assistance and advice on this analysis. 
We thank the extended GLIMPSE team and the IRAC instrument
team members who, together, have helped deliver a genuinely
``legacy'' dataset to the astronomical community.    We
thank the anonymous referee whose astute suggestions
improved this manuscript. This publication makes use of data
products from the Two Micron All Sky Survey, which is a
joint project of the University of Massachusetts and the
Infrared Processing and Analysis Center/California Institute
of Technology, funded by the National Aeronautics and Space
Administration and the National Science Foundation. This
work is based in part on observations made with the Spitzer
Space Telescope, which is operated by the Jet Propulsion
Laboratory, California Institute of Technology under a
contract with NASA, and support was provided by NASA through
awards issued by JPL/Caltech 1368699 and 1367334.

\appendix

\section{The GLIMPSE Pipeline source detection algorithm}

Source detection in the GLIMPSE team's photometry pipeline
utilizes a modified version of the DAOPHOT \citep{Stetson87}
{\tt FIND} routine (See Appendix 2 of the DAOPHOT II User's
Manual; http://www.astro.wisc.edu/glimpse/daophot2.pdf) and is
more fully described in the GLIMPSE documentation on point
source photometry
(http://www.astro.wisc.edu/sirtf/ glimpse\_photometry\_v1.0.pdf).
The GLIMPSE team  modified the DAOPHOT II { \tt FIND}
routine in two ways to accommodate  the high-reliability
requirement of the GLIMPSE Catalogs.  The original DAOPHOT
{\tt FIND} routine determines a global ``sky'' value from the
entire input image. Since the GLIMPSE images contain highly
variable diffuse structure, we first estimate the local sky
level using a 5-pixel median-smoothed version of the input
image.   The ``sky'' value is measured locally in a
5$\times$5 pixel box centered at each pixel location.  In order to
ensure that  detected sources in high-background,
high-structure regions are real, we adopt the maximum sky
value within the box  as the effective sky value.  The
maximum sky value, readout noise, and detector gain are then
used to determine the local noise at each central pixel.
Point source detection is then performed as a two-step
process. Initial detections are conducted using a 3$\sigma$
threshold.  Detected sources are then fit with the
instrument PSF and extracted from the input image. If the
photometric S/N ratio of the extracted source falls below
two, the source is discarded.  Once the initial set of
sources are extracted, the residual image (input image minus
the subtracted sources) is then used as a new input image. 
A new sky image is constructed from this image and secondary
sources are then detected at a 5$\sigma$ threshold.   The
second iteration of {\tt FIND} is performed at a higher
detection level because the initial source subtraction adds
noise around extracted sources, requiring a higher
subsequent detection threshold to avoid  finding false
sources while still detecting real sources hidden in the
wings of the brighter stars.  First-pass source extractions
are then combined with the second-pass extractions to
produce the final list  which we designate as the ``full''
list of candidate sources for each BCD image.  Candidate
sources must still pass additional cross-band and
multi-detection criteria before being included in the
Catalogs or Archives, as described in Section 2 and in
GLIMPSE data  products documents referenced herein.  

{\it Facilities:}  \facility{Spitzer/IRAC (), 2MASS  }

{}

\clearpage

\begin{deluxetable}{lclcccrr}
\rotate
\tabletypesize{\scriptsize}
\tablewidth{9.6in}
\tablecaption{{\it Spitzer} GLIMPSE and Similar Galactic Wide-Area Surveys
 \label{surveys.tab} }
\setlength{\tabcolsep}{0.02in}
\tablehead{
\colhead{Survey} &
\colhead{Coverage} &
\colhead{Approx. Area} &
\colhead{Instrument/Bands} &
\colhead{Exp. Time} &
\colhead{Reference} &
\colhead{Catalog Sources} &
\colhead{Archive Sources}  } 
\startdata
GLIMPSE I   & 10\degr$< | \ell | <$65\degr  ; $| b | <1\degr$                            & 220 sq. deg. & IRAC [3.6],[4.5],[5.6],[8.0] & $2\times1.2$ s     & Churchwell et al. (2009)  & 31,154,438 & 49,133,194 \\
GLIMPSE II  & $| \ell | <10$\degr  ; $|b|$ $\lesssim$1.5\degr\ \tablenotemark{a}         &  54 sq. deg. & IRAC [3.6],[4.5],[5.6],[8.0] & $3\times1.2$ s\tablenotemark{b}     & Churchwell et al. (2009)  & 18,145,818 & 23,125,046 \\
GLIMPSE 3D   & $<|\ell|<$ 31\degr  ;$|b|>1\degr$ \tablenotemark{a}               & 120 sq. deg. & IRAC [3.6],[4.5],[5.6],[8.0] & $3(2)\times1.2$ s\tablenotemark{c} & Churchwell et al. (2009)  & 20,403,915 & 32,214,210 \\
GLIMPSE 360  & $\ell$=65\degr -- 76\degr,82\degr -- 102\degr,109\degr --265\degr         & 511 sq. deg. & IRAC [3.6],[4.5]	       & $3\times10.4$ s    & Whitney et al. (2008)     & 42,602,112 & 49,378,042 \\
             &  $| b | \lesssim3$\degr\tablenotemark{a}                                  &              &                              &                    &                 	        &            &             \\
Deep GLIMPSE &  $\ell$ = 265--350\degr, $b$\degr =-2\degr -- +0.1\degr                     & 208 sq. deg. & IRAC [3.6],[4.5]             & $3\times10.4$ s     & Whitney et al. (2011)	& 38,279,639 & 63,522,165 \\
             &   $\ell$ = 25\degr--65\degr, $b$= 0\degr -- +2.7\degr                       &              &                              &                    &                           &            &             \\
SMOG        & $\ell$=102\degr -- 109\degr ; b= 0 -- 3\degr                               & 21 sq. deg.  & IRAC [3.6],[4.5],[5.6],[8.0] & $4\times10.4$ s    & Carey et al. (2008)       & 2,512,099  &  2,836,618 \\
Cygnus-X    & $\ell$= 76\degr -- 82\degr ;  b = -2.3\degr -- +4.1\degr \tablenotemark{a}  & 24 sq. deg.  & IRAC [3.6],[4.5],[5.6],[8.0] & $3\times10.4$ s    & Hora et al. (2007)        & 3,913,559  &  4,455,066  \\
Vela-Carina & $\ell$=255\degr -- 295\degr ;  b $\approx$ -1.5\degr -- +1.5\degr \tablenotemark{a} & 80 sq. deg.  & IRAC [3.6],[4.5],[5.6],[8.0] & $2\times1.2$ s     & Zasowski et al. (2009)    & 2,000,188  &  4,547,327  \\
\enddata
\tablenotetext{a}{Irregular region; see survey documentation for details.}
\tablenotetext{b}{GLIMPSE~II data products include the $Spitzer$ Galactic Center survey (S. Stolovy; PID=3677) which has five visits.}
\tablenotetext{c}{Some portions of GLIMPSE 3D use two visits and others have three. }
\end{deluxetable}

\begin{deluxetable}{lccccccc}
\tabletypesize{\scriptsize}
\tablecaption{GLIMPSE I/II/3D Catalog Completeness1 as a function of band and
magnitude  \label{comp1cat.tab}}

\setlength{\tabcolsep}{0.02in}
\tablehead{
\colhead{Background } &
\colhead{8--9} &
\colhead{9--10} &
\colhead{10--11} &
\colhead{11--12} &
\colhead{12--13} &
\colhead{13--14} &
\colhead{14--15} \\
\colhead{(MJy/sr)} &
\colhead{(mag.)} &
\colhead{(mag.)} &
\colhead{(mag.)} &
\colhead{(mag.)} &
\colhead{(mag.)} &
\colhead{(mag.)} &
\colhead{(mag.)}
 } 
\tablecolumns{8}
\startdata
\cutinheadc{3.6 $\mu$m}
   1.2--   1.8 &  0.98 &  1.00 &  0.99 &  1.00 &  0.98 &  0.89 &  0.26 \\
   1.8--   2.6 &  0.99 &  1.00 &  1.00 &  0.99 &  1.00 &  0.96 &  0.23 \\
   2.6--   3.8 &  0.99 &  0.99 &  0.99 &  0.99 &  0.98 &  0.95 &  0.24 \\
   3.8--   5.6 &  0.99 &  0.99 &  0.99 &  0.98 &  0.96 &  0.91 &  0.22 \\
   5.6--   8.2 &  0.99 &  0.99 &  0.98 &  0.95 &  0.92 &  0.81 &  0.18 \\
   8.2--  12.0 &  0.97 &  0.97 &  0.96 &  0.91 &  0.81 &  0.57 &  0.10 \\
  12.0--  17.6 &  0.96 &  0.96 &  0.92 &  0.84 &  0.64 &  0.30 &  0.03 \\
  17.6--  25.9 &  0.95 &  0.91 &  0.84 &  0.68 &  0.43 &  0.20 &  0.02 \\
  25.9--  37.9 &  0.87 &  0.77 &  0.70 &  0.44 &  0.29 &  0.06 &  0.00 \\
  37.9--  55.7 &  0.68 &  0.58 &  0.29 &  0.13 &  0.02 &  0.00 &  0.02 \\
  55.7--  81.8 &  0.64 &  0.41 &  0.20 &  0.12 &  0.00 &  0.00 &  0.00 \\
  81.8-- 120.0 &  0.31 &  0.07 &  0.00 &  0.06 &  0.00 &  0.00 &  0.00 \\
\hline
\cutinheadc{4.5 $\mu$m}
   0.8--   1.1 &  0.98 &  0.99 &  1.00 &  0.99 &  1.00 &  0.93 &  0.24 \\
   1.1--   1.6 &  0.99 &  1.00 &  1.00 &  0.99 &  0.99 &  0.96 &  0.23 \\
   1.6--   2.4 &  0.99 &  1.00 &  0.99 &  0.99 &  0.98 &  0.95 &  0.25 \\
   2.4--   3.4 &  0.99 &  0.99 &  1.00 &  0.99 &  0.98 &  0.93 &  0.23 \\
   3.4--   4.8 &  0.99 &  0.99 &  0.99 &  0.98 &  0.95 &  0.90 &  0.20 \\
   4.8--   6.9 &  0.99 &  0.99 &  0.98 &  0.94 &  0.92 &  0.80 &  0.18 \\
   6.9--   9.9 &  0.98 &  0.97 &  0.97 &  0.92 &  0.86 &  0.63 &  0.10 \\
   9.9--  14.2 &  0.96 &  0.96 &  0.92 &  0.87 &  0.68 &  0.34 &  0.04 \\
  14.2--  20.4 &  0.94 &  0.92 &  0.88 &  0.74 &  0.54 &  0.26 &  0.01 \\
  20.4--  29.2 &  0.97 &  0.87 &  0.78 &  0.64 &  0.36 &  0.12 &  0.01 \\
  29.2--  41.9 &  0.76 &  0.68 &  0.61 &  0.32 &  0.17 &  0.02 &  0.00 \\
  41.9--  60.0 &  0.67 &  0.59 &  0.31 &  0.20 &  0.02 &  0.00 &  0.02 \\
\hline
\cutinheadc{5.8 $\mu$m}
   9.0--  12.9 &  0.99 &  1.00 &  1.00 &  0.99 &  0.99 &  0.95 &  0.25 \\
  12.9--  18.6 &  0.99 &  0.99 &  0.99 &  0.99 &  0.97 &  0.93 &  0.25 \\
  18.6--  26.7 &  0.99 &  0.99 &  0.99 &  0.98 &  0.96 &  0.91 &  0.22 \\
  26.7--  38.4 &  0.99 &  0.99 &  0.99 &  0.96 &  0.94 &  0.88 &  0.20 \\
  38.4--  55.2 &  0.98 &  0.98 &  0.97 &  0.94 &  0.93 &  0.83 &  0.18 \\
  55.2--  79.4 &  0.98 &  0.96 &  0.95 &  0.93 &  0.89 &  0.72 &  0.14 \\
  79.4-- 114.1 &  0.97 &  0.96 &  0.97 &  0.94 &  0.83 &  0.47 &  0.06 \\
 114.1-- 164.0 &  0.96 &  0.97 &  0.94 &  0.86 &  0.66 &  0.29 &  0.03 \\
 164.0-- 235.7 &  0.94 &  0.93 &  0.93 &  0.76 &  0.49 &  0.23 &  0.02 \\
 235.7-- 338.8 &  0.93 &  0.88 &  0.68 &  0.47 &  0.27 &  0.02 &  0.00 \\
 338.8-- 487.0 &  0.75 &  0.58 &  0.33 &  0.20 &  0.02 &  0.00 &  0.02 \\
 487.0-- 700.0 &  0.77 &  0.66 &  0.16 &  0.09 &  0.00 &  0.00 &  0.00 \\
\hline
\cutinheadc{8.0 $\mu$m}
  25.0--  35.2 &  0.99 &  1.00 &  1.00 &  0.98 &  0.98 &  0.92 &  0.26 \\
  35.2--  49.5 &  0.99 &  0.99 &  0.99 &  0.99 &  0.95 &  0.92 &  0.24 \\
  49.5--  69.6 &  0.98 &  0.99 &  0.99 &  0.98 &  0.96 &  0.91 &  0.23 \\
  69.6--  97.9 &  0.98 &  0.99 &  0.99 &  0.97 &  0.94 &  0.89 &  0.21 \\
  97.9-- 137.7 &  0.99 &  0.99 &  0.98 &  0.96 &  0.95 &  0.87 &  0.19 \\
 137.7-- 193.6 &  0.99 &  0.98 &  0.97 &  0.95 &  0.93 &  0.84 &  0.18 \\
 193.6-- 272.4 &  0.99 &  0.98 &  0.97 &  0.96 &  0.91 &  0.71 &  0.12 \\
 272.4-- 383.2 &  0.98 &  0.98 &  0.97 &  0.94 &  0.82 &  0.45 &  0.06 \\
 383.2-- 539.0 &  0.98 &  0.98 &  0.95 &  0.86 &  0.65 &  0.29 &  0.02 \\
 539.0-- 758.1 &  0.95 &  0.97 &  0.93 &  0.71 &  0.46 &  0.19 &  0.03 \\
 758.1--1066.4 &  0.90 &  0.82 &  0.64 &  0.45 &  0.25 &  0.01 &  0.00 \\
1066.4--1500.0 &  0.86 &  0.56 &  0.25 &  0.16 &  0.00 &  0.00 &  0.02 \\
\enddata
\end{deluxetable}

\begin{deluxetable}{lccccccc}
\tabletypesize{\scriptsize}
\tablecaption{GLIMPSE I/II/3D Catalog Completeness2 as a function of band and
magnitude  \label{comp2cat.tab}}

\setlength{\tabcolsep}{0.02in}
\tablehead{
\colhead{Background } &
\colhead{8--9} &
\colhead{9--10} &
\colhead{10--11} &
\colhead{11--12} &
\colhead{12--13} &
\colhead{13--14} &
\colhead{14--15} \\
\colhead{(MJy/sr)} &
\colhead{(mag.)} &
\colhead{(mag.)} &
\colhead{(mag.)} &
\colhead{(mag.)} &
\colhead{(mag.)} &
\colhead{(mag.)} &
\colhead{(mag.)}
 } 
\tablecolumns{8}
\startdata
\cutinheadc{3.6 $\mu$m}
   1.2--   1.8 &  0.98 &  1.00 &  0.99 &  1.00 &  0.98 &  0.89 &  0.26 \\
   1.8--   2.6 &  0.99 &  1.00 &  0.99 &  0.99 &  1.00 &  0.96 &  0.23 \\
   2.6--   3.8 &  0.98 &  0.99 &  0.99 &  0.99 &  0.98 &  0.95 &  0.24 \\
   3.8--   5.6 &  0.97 &  0.98 &  0.98 &  0.97 &  0.96 &  0.91 &  0.22 \\
   5.6--   8.2 &  0.96 &  0.96 &  0.96 &  0.94 &  0.92 &  0.81 &  0.18 \\
   8.2--  12.0 &  0.93 &  0.93 &  0.92 &  0.90 &  0.81 &  0.57 &  0.10 \\
  12.0--  17.6 &  0.92 &  0.90 &  0.89 &  0.83 &  0.64 &  0.30 &  0.03 \\
  17.6--  25.9 &  0.90 &  0.88 &  0.81 &  0.66 &  0.43 &  0.20 &  0.02 \\
  25.9--  37.9 &  0.80 &  0.73 &  0.65 &  0.42 &  0.29 &  0.06 &  0.00 \\
  37.9--  55.7 &  0.58 &  0.53 &  0.29 &  0.11 &  0.02 &  0.00 &  0.02 \\
  55.7--  81.8 &  0.64 &  0.41 &  0.20 &  0.12 &  0.00 &  0.00 &  0.00 \\
  81.8-- 120.0 &  0.25 &  0.07 &  0.00 &  0.06 &  0.00 &  0.00 &  0.00 \\
\hline
\cutinheadc{4.5 $\mu$m}
   0.8--   1.1 &  0.96 &  0.99 &  0.98 &  0.99 &  1.00 &  0.93 &  0.24 \\
   1.1--   1.6 &  0.99 &  0.99 &  0.99 &  0.99 &  0.99 &  0.96 &  0.23 \\
   1.6--   2.4 &  0.98 &  0.98 &  0.98 &  0.98 &  0.98 &  0.95 &  0.25 \\
   2.4--   3.4 &  0.97 &  0.98 &  0.98 &  0.98 &  0.98 &  0.93 &  0.23 \\
   3.4--   4.8 &  0.95 &  0.96 &  0.95 &  0.96 &  0.95 &  0.90 &  0.20 \\
   4.8--   6.9 &  0.93 &  0.92 &  0.93 &  0.93 &  0.92 &  0.80 &  0.18 \\
   6.9--   9.9 &  0.90 &  0.90 &  0.93 &  0.91 &  0.86 &  0.63 &  0.10 \\
   9.9--  14.2 &  0.87 &  0.88 &  0.87 &  0.86 &  0.68 &  0.34 &  0.04 \\
  14.2--  20.4 &  0.87 &  0.88 &  0.84 &  0.73 &  0.54 &  0.26 &  0.01 \\
  20.4--  29.2 &  0.90 &  0.83 &  0.77 &  0.63 &  0.36 &  0.12 &  0.01 \\
  29.2--  41.9 &  0.70 &  0.65 &  0.61 &  0.32 &  0.17 &  0.02 &  0.00 \\
  41.9--  60.0 &  0.64 &  0.59 &  0.29 &  0.20 &  0.02 &  0.00 &  0.02 \\
\hline
\cutinheadc{5.8 $\mu$m}
   9.0--  12.9 &  0.99 &  1.00 &  1.00 &  0.99 &  0.64 &  0.05 &  0.00 \\
  12.9--  18.6 &  0.99 &  0.99 &  0.99 &  0.98 &  0.53 &  0.03 &  0.00 \\
  18.6--  26.7 &  0.99 &  0.99 &  0.99 &  0.97 &  0.42 &  0.02 &  0.00 \\
  26.7--  38.4 &  0.99 &  0.99 &  0.98 &  0.92 &  0.29 &  0.01 &  0.00 \\
  38.4--  55.2 &  0.98 &  0.98 &  0.96 &  0.81 &  0.18 &  0.01 &  0.00 \\
  55.2--  79.4 &  0.98 &  0.96 &  0.89 &  0.56 &  0.06 &  0.01 &  0.00 \\
  79.4-- 114.1 &  0.97 &  0.91 &  0.77 &  0.27 &  0.02 &  0.01 &  0.00 \\
 114.1-- 164.0 &  0.92 &  0.84 &  0.49 &  0.11 &  0.01 &  0.00 &  0.00 \\
 164.0-- 235.7 &  0.83 &  0.65 &  0.35 &  0.05 &  0.01 &  0.00 &  0.00 \\
 235.7-- 338.8 &  0.64 &  0.29 &  0.05 &  0.03 &  0.01 &  0.00 &  0.00 \\
 338.8-- 487.0 &  0.22 &  0.10 &  0.05 &  0.00 &  0.00 &  0.00 &  0.00 \\
 487.0-- 700.0 &  0.13 &  0.09 &  0.00 &  0.00 &  0.00 &  0.00 &  0.00 \\
\hline
\cutinheadc{8.0 $\mu$m}
  25.0--  35.2 &  0.99 &  1.00 &  0.99 &  0.92 &  0.24 &  0.00 &  0.00 \\
  35.2--  49.5 &  0.98 &  0.99 &  0.99 &  0.87 &  0.15 &  0.00 &  0.00 \\
  49.5--  69.6 &  0.98 &  0.99 &  0.98 &  0.69 &  0.05 &  0.00 &  0.00 \\
  69.6--  97.9 &  0.98 &  0.98 &  0.95 &  0.50 &  0.02 &  0.00 &  0.00 \\
  97.9-- 137.7 &  0.98 &  0.97 &  0.83 &  0.24 &  0.01 &  0.00 &  0.00 \\
 137.7-- 193.6 &  0.97 &  0.90 &  0.57 &  0.07 &  0.00 &  0.00 &  0.00 \\
 193.6-- 272.4 &  0.89 &  0.66 &  0.21 &  0.01 &  0.00 &  0.00 &  0.00 \\
 272.4-- 383.2 &  0.67 &  0.30 &  0.04 &  0.00 &  0.00 &  0.00 &  0.00 \\
 383.2-- 539.0 &  0.38 &  0.16 &  0.01 &  0.00 &  0.00 &  0.00 &  0.00 \\
 539.0-- 758.1 &  0.20 &  0.03 &  0.01 &  0.01 &  0.00 &  0.00 &  0.00 \\
 758.1--1066.4 &  0.04 &  0.02 &  0.00 &  0.00 &  0.00 &  0.00 &  0.00 \\
1066.4--1500.0 &  0.00 &  0.00 &  0.00 &  0.00 &  0.00 &  0.00 &  0.00 \\
\enddata
\end{deluxetable}

\begin{deluxetable}{lccccccc}
\tabletypesize{\scriptsize}
\tablecaption{GLIMPSE I/II/3D Archive Completeness1 as a function of band and
magnitude  \label{comp1arc.tab}}

\setlength{\tabcolsep}{0.02in}
\tablehead{
\colhead{Background } &
\colhead{8--9} &
\colhead{9--10} &
\colhead{10--11} &
\colhead{11--12} &
\colhead{12--13} &
\colhead{13--14} &
\colhead{14--15} \\
\colhead{(MJy/sr)} &
\colhead{(mag.)} &
\colhead{(mag.)} &
\colhead{(mag.)} &
\colhead{(mag.)} &
\colhead{(mag.)} &
\colhead{(mag.)} &
\colhead{(mag.)}
 } 
\tablecolumns{8}
\startdata
\cutinheadc{3.6 $\mu$m}
   1.2--   1.8 &  1.00 &  1.00 &  1.00 &  1.00 &  1.00 &  0.96 &  0.68 \\
   1.8--   2.6 &  1.00 &  1.00 &  1.00 &  1.00 &  1.00 &  0.99 &  0.72 \\
   2.6--   3.8 &  1.00 &  1.00 &  1.00 &  1.00 &  1.00 &  0.99 &  0.68 \\
   3.8--   5.6 &  1.00 &  1.00 &  1.00 &  1.00 &  1.00 &  0.98 &  0.59 \\
   5.6--   8.2 &  1.00 &  1.00 &  1.00 &  1.00 &  0.99 &  0.92 &  0.44 \\
   8.2--  12.0 &  1.00 &  1.00 &  1.00 &  0.99 &  0.94 &  0.72 &  0.21 \\
  12.0--  17.6 &  1.00 &  1.00 &  0.99 &  0.96 &  0.81 &  0.44 &  0.07 \\
  17.6--  25.9 &  0.99 &  0.98 &  0.94 &  0.78 &  0.56 &  0.30 &  0.04 \\
  25.9--  37.9 &  0.99 &  0.93 &  0.82 &  0.56 &  0.41 &  0.10 &  0.00 \\
  37.9--  55.7 &  0.87 &  0.84 &  0.46 &  0.24 &  0.13 &  0.01 &  0.02 \\
  55.7--  81.8 &  0.81 &  0.59 &  0.33 &  0.14 &  0.06 &  0.00 &  0.00 \\
  81.8-- 120.0 &  0.44 &  0.33 &  0.07 &  0.06 &  0.00 &  0.00 &  0.00 \\
\hline
\cutinheadc{4.5 $\mu$m}
   0.8--   1.1 &  0.99 &  1.00 &  1.00 &  1.00 &  1.00 &  0.99 &  0.75 \\
   1.1--   1.6 &  1.00 &  1.00 &  1.00 &  1.00 &  1.00 &  0.99 &  0.73 \\
   1.6--   2.4 &  1.00 &  1.00 &  1.00 &  1.00 &  1.00 &  0.99 &  0.69 \\
   2.4--   3.4 &  1.00 &  1.00 &  1.00 &  1.00 &  1.00 &  0.98 &  0.64 \\
   3.4--   4.8 &  1.00 &  1.00 &  1.00 &  1.00 &  1.00 &  0.97 &  0.55 \\
   4.8--   6.9 &  1.00 &  1.00 &  1.00 &  1.00 &  0.99 &  0.91 &  0.44 \\
   6.9--   9.9 &  1.00 &  1.00 &  1.00 &  0.99 &  0.95 &  0.78 &  0.24 \\
   9.9--  14.2 &  1.00 &  1.00 &  0.99 &  0.96 &  0.84 &  0.48 &  0.09 \\
  14.2--  20.4 &  0.99 &  0.98 &  0.98 &  0.85 &  0.68 &  0.39 &  0.04 \\
  20.4--  29.2 &  0.99 &  0.98 &  0.88 &  0.71 &  0.51 &  0.18 &  0.02 \\
  29.2--  41.9 &  0.95 &  0.86 &  0.72 &  0.49 &  0.26 &  0.06 &  0.00 \\
  41.9--  60.0 &  0.88 &  0.76 &  0.53 &  0.27 &  0.13 &  0.00 &  0.02 \\
\hline
\cutinheadc{5.8 $\mu$m}
   9.0--  12.9 &  1.00 &  1.00 &  1.00 &  1.00 &  1.00 &  0.98 &  0.72 \\
  12.9--  18.6 &  1.00 &  1.00 &  1.00 &  1.00 &  1.00 &  0.98 &  0.67 \\
  18.6--  26.7 &  1.00 &  1.00 &  1.00 &  1.00 &  0.99 &  0.97 &  0.61 \\
  26.7--  38.4 &  1.00 &  1.00 &  1.00 &  1.00 &  0.99 &  0.95 &  0.56 \\
  38.4--  55.2 &  1.00 &  1.00 &  1.00 &  0.99 &  0.98 &  0.92 &  0.48 \\
  55.2--  79.4 &  1.00 &  0.99 &  0.99 &  0.98 &  0.95 &  0.83 &  0.31 \\
  79.4-- 114.1 &  1.00 &  0.99 &  0.99 &  0.97 &  0.92 &  0.61 &  0.13 \\
 114.1-- 164.0 &  0.99 &  0.99 &  0.98 &  0.92 &  0.80 &  0.45 &  0.07 \\
 164.0-- 235.7 &  0.99 &  0.95 &  0.94 &  0.82 &  0.62 &  0.31 &  0.03 \\
 235.7-- 338.8 &  0.97 &  0.94 &  0.79 &  0.60 &  0.38 &  0.05 &  0.00 \\
 338.8-- 487.0 &  0.86 &  0.71 &  0.51 &  0.33 &  0.12 &  0.03 &  0.02 \\
 487.0-- 700.0 &  0.87 &  0.83 &  0.30 &  0.12 &  0.09 &  0.00 &  0.00 \\
\hline
\cutinheadc{8.0 $\mu$m}
  25.0--  35.2 &  1.00 &  1.00 &  1.00 &  1.00 &  1.00 &  0.98 &  0.72 \\
  35.2--  49.5 &  1.00 &  1.00 &  1.00 &  1.00 &  0.99 &  0.97 &  0.66 \\
  49.5--  69.6 &  1.00 &  1.00 &  1.00 &  1.00 &  0.99 &  0.97 &  0.64 \\
  69.6--  97.9 &  1.00 &  1.00 &  1.00 &  0.99 &  0.99 &  0.96 &  0.59 \\
  97.9-- 137.7 &  1.00 &  1.00 &  1.00 &  0.99 &  0.98 &  0.95 &  0.54 \\
 137.7-- 193.6 &  1.00 &  1.00 &  1.00 &  0.99 &  0.98 &  0.92 &  0.45 \\
 193.6-- 272.4 &  1.00 &  0.99 &  1.00 &  0.99 &  0.97 &  0.82 &  0.28 \\
 272.4-- 383.2 &  1.00 &  1.00 &  0.99 &  0.98 &  0.93 &  0.61 &  0.12 \\
 383.2-- 539.0 &  1.00 &  0.99 &  0.97 &  0.93 &  0.78 &  0.43 &  0.05 \\
 539.0-- 758.1 &  1.00 &  0.98 &  0.96 &  0.79 &  0.58 &  0.27 &  0.05 \\
 758.1--1066.4 &  0.98 &  0.92 &  0.81 &  0.60 &  0.36 &  0.03 &  0.00 \\
1066.4--1500.0 &  0.92 &  0.71 &  0.44 &  0.24 &  0.17 &  0.02 &  0.02 \\
\enddata
\end{deluxetable}

\begin{deluxetable}{lccccccc}
\tabletypesize{\scriptsize}
\tablecaption{GLIMPSE I/II/3D Archive Completeness2 as a function of band and
magnitude  \label{comp2arc.tab}}

\setlength{\tabcolsep}{0.02in}
\tablehead{
\colhead{Background } &
\colhead{8--9} &
\colhead{9--10} &
\colhead{10--11} &
\colhead{11--12} &
\colhead{12--13} &
\colhead{13--14} &
\colhead{14--15} \\
\colhead{(MJy/sr)} &
\colhead{(mag.)} &
\colhead{(mag.)} &
\colhead{(mag.)} &
\colhead{(mag.)} &
\colhead{(mag.)} &
\colhead{(mag.)} &
\colhead{(mag.)}
 } 
\tablecolumns{8}
\startdata
\cutinheadc{3.6 $\mu$m}
   1.2--   1.8 &  0.98 &  1.00 &  1.00 &  1.00 &  1.00 &  0.96 &  0.68 \\
   1.8--   2.6 &  0.99 &  1.00 &  1.00 &  1.00 &  1.00 &  0.99 &  0.72 \\
   2.6--   3.8 &  0.99 &  1.00 &  1.00 &  1.00 &  1.00 &  0.99 &  0.67 \\
   3.8--   5.6 &  1.00 &  1.00 &  1.00 &  1.00 &  1.00 &  0.98 &  0.59 \\
   5.6--   8.2 &  1.00 &  1.00 &  1.00 &  1.00 &  0.99 &  0.92 &  0.44 \\
   8.2--  12.0 &  1.00 &  1.00 &  1.00 &  0.98 &  0.94 &  0.71 &  0.21 \\
  12.0--  17.6 &  0.99 &  0.99 &  0.97 &  0.95 &  0.80 &  0.44 &  0.07 \\
  17.6--  25.9 &  0.98 &  0.97 &  0.91 &  0.75 &  0.55 &  0.30 &  0.04 \\
  25.9--  37.9 &  0.90 &  0.88 &  0.76 &  0.53 &  0.40 &  0.10 &  0.00 \\
  37.9--  55.7 &  0.70 &  0.77 &  0.41 &  0.20 &  0.11 &  0.01 &  0.02 \\
  55.7--  81.8 &  0.79 &  0.51 &  0.33 &  0.14 &  0.06 &  0.00 &  0.00 \\
  81.8-- 120.0 &  0.38 &  0.20 &  0.07 &  0.06 &  0.00 &  0.00 &  0.00 \\
\hline
\cutinheadc{4.5 $\mu$m}
   0.8--   1.1 &  0.98 &  0.99 &  1.00 &  1.00 &  1.00 &  0.98 &  0.49 \\
   1.1--   1.6 &  1.00 &  1.00 &  1.00 &  1.00 &  1.00 &  0.97 &  0.51 \\
   1.6--   2.4 &  0.99 &  1.00 &  1.00 &  1.00 &  1.00 &  0.98 &  0.50 \\
   2.4--   3.4 &  1.00 &  1.00 &  1.00 &  1.00 &  1.00 &  0.96 &  0.44 \\
   3.4--   4.8 &  1.00 &  1.00 &  1.00 &  1.00 &  0.99 &  0.94 &  0.36 \\
   4.8--   6.9 &  1.00 &  1.00 &  1.00 &  0.99 &  0.98 &  0.86 &  0.28 \\
   6.9--   9.9 &  1.00 &  0.99 &  0.99 &  0.98 &  0.92 &  0.67 &  0.14 \\
   9.9--  14.2 &  0.98 &  0.98 &  0.95 &  0.92 &  0.76 &  0.37 &  0.05 \\
  14.2--  20.4 &  0.98 &  0.96 &  0.93 &  0.80 &  0.59 &  0.28 &  0.01 \\
  20.4--  29.2 &  0.94 &  0.96 &  0.82 &  0.66 &  0.38 &  0.14 &  0.01 \\
  29.2--  41.9 &  0.85 &  0.79 &  0.66 &  0.38 &  0.17 &  0.02 &  0.00 \\
  41.9--  60.0 &  0.81 &  0.67 &  0.36 &  0.21 &  0.04 &  0.00 &  0.02 \\
\hline
\cutinheadc{5.8 $\mu$m}
   9.0--  12.9 &  1.00 &  1.00 &  1.00 &  1.00 &  0.64 &  0.05 &  0.00 \\
  12.9--  18.6 &  0.99 &  1.00 &  1.00 &  0.99 &  0.55 &  0.03 &  0.00 \\
  18.6--  26.7 &  1.00 &  1.00 &  1.00 &  0.98 &  0.43 &  0.02 &  0.00 \\
  26.7--  38.4 &  0.99 &  1.00 &  0.99 &  0.94 &  0.31 &  0.01 &  0.00 \\
  38.4--  55.2 &  1.00 &  1.00 &  0.98 &  0.84 &  0.18 &  0.01 &  0.00 \\
  55.2--  79.4 &  0.99 &  0.98 &  0.93 &  0.59 &  0.06 &  0.01 &  0.00 \\
  79.4-- 114.1 &  0.98 &  0.93 &  0.79 &  0.27 &  0.02 &  0.01 &  0.00 \\
 114.1-- 164.0 &  0.94 &  0.85 &  0.49 &  0.11 &  0.01 &  0.01 &  0.00 \\
 164.0-- 235.7 &  0.88 &  0.66 &  0.35 &  0.05 &  0.01 &  0.01 &  0.00 \\
 235.7-- 338.8 &  0.67 &  0.29 &  0.06 &  0.04 &  0.01 &  0.00 &  0.00 \\
 338.8-- 487.0 &  0.25 &  0.12 &  0.07 &  0.00 &  0.00 &  0.00 &  0.00 \\
 487.0-- 700.0 &  0.16 &  0.09 &  0.00 &  0.00 &  0.00 &  0.00 &  0.00 \\
\hline
\cutinheadc{8.0 $\mu$m}
  25.0--  35.2 &  1.00 &  1.00 &  1.00 &  0.94 &  0.24 &  0.00 &  0.00 \\
  35.2--  49.5 &  0.99 &  1.00 &  1.00 &  0.87 &  0.15 &  0.01 &  0.00 \\
  49.5--  69.6 &  1.00 &  0.99 &  0.99 &  0.69 &  0.05 &  0.00 &  0.00 \\
  69.6--  97.9 &  0.99 &  0.99 &  0.96 &  0.50 &  0.02 &  0.00 &  0.00 \\
  97.9-- 137.7 &  0.99 &  0.98 &  0.84 &  0.24 &  0.01 &  0.00 &  0.00 \\
 137.7-- 193.6 &  0.98 &  0.91 &  0.57 &  0.07 &  0.00 &  0.00 &  0.00 \\
 193.6-- 272.4 &  0.90 &  0.66 &  0.21 &  0.01 &  0.00 &  0.00 &  0.00 \\
 272.4-- 383.2 &  0.67 &  0.30 &  0.04 &  0.00 &  0.00 &  0.00 &  0.00 \\
 383.2-- 539.0 &  0.40 &  0.16 &  0.01 &  0.00 &  0.00 &  0.00 &  0.00 \\
 539.0-- 758.1 &  0.21 &  0.03 &  0.01 &  0.01 &  0.00 &  0.00 &  0.00 \\
 758.1--1066.4 &  0.04 &  0.02 &  0.00 &  0.00 &  0.00 &  0.00 &  0.00 \\
1066.4--1500.0 &  0.00 &  0.00 &  0.00 &  0.00 &  0.00 &  0.00 &  0.00 \\
\enddata
\end{deluxetable}

\begin{deluxetable}{lcccccccc}
\tabletypesize{\scriptsize}
\tablecaption{GLIMPSE 360 \& Deep GLIMPSE Catalog Completeness1\tablenotemark{a} as a function of band and
magnitude  \label{G360C1.tab}}
\setlength{\tabcolsep}{0.02in}
\tablehead{
\colhead{Background } &
\colhead{11--12} &
\colhead{12--13} &
\colhead{13--14} &
\colhead{14--15} &
\colhead{15--16} &
\colhead{16--17} &
\colhead{17--18} \\
\colhead{(MJy/sr)} &
\colhead{(mag.)} &
\colhead{(mag.)} &
\colhead{(mag.)} &
\colhead{(mag.)} &
\colhead{(mag.)} &
\colhead{(mag.)} &
\colhead{(mag.)} } 
\tablecolumns{8}
\startdata
\cutinheadc{3.6$\mu$m}
  0.12--  0.21 &  0.97 &  0.97 &  0.97 &  0.97 &  0.97 &  0.85 &  0.11 \\
  0.21--  0.37 &  0.95 &  0.94 &  0.95 &  0.95 &  0.95 &  0.80 &  0.09 \\
  0.37--  0.66 &  0.92 &  0.93 &  0.93 &  0.92 &  0.92 &  0.70 &  0.05 \\
  0.66--  1.17 &  0.90 &  0.90 &  0.90 &  0.90 &  0.88 &  0.56 &  0.02 \\
  1.17--  2.06 &  0.88 &  0.88 &  0.89 &  0.87 &  0.82 &  0.36 &  0.01 \\
  2.06--  3.63 &  0.87 &  0.89 &  0.89 &  0.84 &  0.65 &  0.14 &  0.00 \\
  3.63--  6.41 &  0.85 &  0.82 &  0.75 &  0.63 &  0.29 &  0.02 &  0.00 \\
  6.41-- 11.32 &  0.71 &  0.62 &  0.42 &  0.24 &  0.04 &  0.00 &  0.00 \\
 11.32-- 19.99 &  0.43 &  0.31 &  0.10 &  0.03 &  0.00 &  0.00 &  0.00 \\
 19.99-- 35.29 &  0.11 &  0.03 &  0.02 &  0.00 &  0.00 &  0.00 &  0.00 \\
 35.29-- 62.31 &  0.00 &  0.00 &  0.00 &  0.00 &  0.00 &  0.00 &  0.00 \\
 62.31--110.00 &  0.00 &  0.00 &  0.00 &  0.00 &  0.00 &  0.00 &  0.00 \\
\hline
\cutinheadc{4.5 $\mu$m}
  0.07--  0.12 &  0.98 &  0.98 &  0.97 &  0.98 &  0.98 &  0.86 &  0.11 \\
  0.12--  0.22 &  0.95 &  0.96 &  0.96 &  0.96 &  0.96 &  0.83 &  0.10 \\
  0.22--  0.38 &  0.93 &  0.93 &  0.93 &  0.93 &  0.93 &  0.75 &  0.07 \\
  0.38--  0.66 &  0.92 &  0.92 &  0.92 &  0.92 &  0.91 &  0.65 &  0.04 \\
  0.66--  1.17 &  0.89 &  0.89 &  0.89 &  0.88 &  0.85 &  0.49 &  0.01 \\
  1.17--  2.05 &  0.87 &  0.89 &  0.89 &  0.86 &  0.78 &  0.28 &  0.00 \\
  2.05--  3.60 &  0.88 &  0.88 &  0.86 &  0.82 &  0.56 &  0.08 &  0.00 \\
  3.60--  6.32 &  0.82 &  0.80 &  0.71 &  0.54 &  0.21 &  0.01 &  0.00 \\
  6.32-- 11.09 &  0.72 &  0.61 &  0.42 &  0.26 &  0.03 &  0.00 &  0.00 \\
 11.09-- 19.47 &  0.45 &  0.27 &  0.09 &  0.02 &  0.00 &  0.00 &  0.00 \\
 19.47-- 34.18 &  0.14 &  0.07 &  0.03 &  0.00 &  0.00 &  0.00 &  0.00 \\
 34.18-- 60.00 &  0.00 &  0.00 &  0.00 &  0.00 &  0.00 &  0.00 &  0.00 \\
\enddata
\tablenotetext{a}{Completeness of the GLIMPSE~360/Deep GLIMPSE Catalogs
is significantly reduced in some locations by the density of saturated stars and their
extended PSF wings (R=24 pixels).  Saturated star density varies greatly
with Galactic location, so these data should be regarded as illustrative of general trends
rather than definitive for all regions.  Users requiring a precise estimate of
completeness in the GLIMPSE~360/Deep GLIMPSE Catalogs should conduct their
own analysis of saturated star density specific to their
region of interest and its impact on Catalog completeness.  We recommend
the  GLIMPSE~360/Deep GLIMPSE Archive in lieu of the Catalog for applications requiring 
high completeness.}
\end{deluxetable}

\begin{deluxetable}{lccccccc}
\tabletypesize{\scriptsize}
\tablecaption{GLIMPSE 360  \& Deep GLIMPSE  Catalog Completeness2\tablenotemark{a} as a function of band and
magnitude  \label{G360C2.tab}}
\setlength{\tabcolsep}{0.02in}
\tablehead{
\colhead{Background } &
\colhead{11--12} &
\colhead{12--13} &
\colhead{13--14} &
\colhead{14--15} &
\colhead{15--16} &
\colhead{16--17} &
\colhead{17--18} \\
\colhead{(MJy/sr)} &
\colhead{(mag.)} &
\colhead{(mag.)} &
\colhead{(mag.)} &
\colhead{(mag.)} &
\colhead{(mag.)} &
\colhead{(mag.)} &
\colhead{(mag.)}
 } 
\tablecolumns{8}
\startdata
\cutinheadc{3.6 $\mu$m}
  0.12--  0.21 &  0.96 &  0.96 &  0.95 &  0.96 &  0.96 &  0.83 &  0.10 \\
  0.21--  0.37 &  0.93 &  0.93 &  0.93 &  0.93 &  0.93 &  0.78 &  0.08 \\
  0.37--  0.66 &  0.91 &  0.91 &  0.91 &  0.91 &  0.90 &  0.68 &  0.04 \\
  0.66--  1.17 &  0.89 &  0.89 &  0.89 &  0.88 &  0.86 &  0.54 &  0.02 \\
  1.17--  2.06 &  0.87 &  0.87 &  0.88 &  0.87 &  0.80 &  0.34 &  0.01 \\
  2.06--  3.63 &  0.86 &  0.88 &  0.88 &  0.83 &  0.62 &  0.12 &  0.00 \\
  3.63--  6.41 &  0.84 &  0.80 &  0.73 &  0.60 &  0.26 &  0.02 &  0.00 \\
  6.41-- 11.32 &  0.69 &  0.60 &  0.41 &  0.23 &  0.03 &  0.00 &  0.00 \\
 11.32-- 19.99 &  0.42 &  0.28 &  0.08 &  0.03 &  0.00 &  0.00 &  0.00 \\
 19.99-- 35.29 &  0.11 &  0.03 &  0.02 &  0.00 &  0.00 &  0.00 &  0.00 \\
 35.29-- 62.31 &  0.00 &  0.00 &  0.00 &  0.00 &  0.00 &  0.00 &  0.00 \\
 62.31--110.00 &  0.00 &  0.00 &  0.00 &  0.00 &  0.00 &  0.00 &  0.00 \\
\hline
\cutinheadc{4.5 $\mu$m}
  0.07--  0.12 &  0.96 &  0.96 &  0.96 &  0.96 &  0.96 &  0.67 &  0.04 \\
  0.12--  0.22 &  0.94 &  0.94 &  0.95 &  0.95 &  0.94 &  0.63 &  0.03 \\
  0.22--  0.38 &  0.92 &  0.91 &  0.92 &  0.91 &  0.91 &  0.56 &  0.02 \\
  0.38--  0.66 &  0.91 &  0.91 &  0.91 &  0.91 &  0.89 &  0.46 &  0.01 \\
  0.66--  1.17 &  0.88 &  0.88 &  0.88 &  0.87 &  0.82 &  0.31 &  0.00 \\
  1.17--  2.05 &  0.86 &  0.88 &  0.88 &  0.85 &  0.69 &  0.16 &  0.00 \\
  2.05--  3.60 &  0.87 &  0.87 &  0.85 &  0.77 &  0.43 &  0.04 &  0.00 \\
  3.60--  6.32 &  0.81 &  0.78 &  0.67 &  0.46 &  0.14 &  0.00 &  0.00 \\
  6.32-- 11.09 &  0.70 &  0.56 &  0.35 &  0.19 &  0.02 &  0.00 &  0.00 \\
 11.09-- 19.47 &  0.42 &  0.23 &  0.07 &  0.02 &  0.00 &  0.00 &  0.00 \\
 19.47-- 34.18 &  0.12 &  0.03 &  0.02 &  0.00 &  0.00 &  0.00 &  0.00 \\
 34.18-- 60.00 &  0.00 &  0.00 &  0.00 &  0.00 &  0.00 &  0.00 &  0.00 \\
\enddata
\tablenotetext{a}{Completeness of the GLIMPSE~360/Deep GLIMPSE Catalogs
is significantly reduced in some locations by the density of saturated stars and their
extended PSF wings (R=24 pixels).  Saturated star density varies greatly
with Galactic location, so these data should be regarded as illustrative of general trends
rather than definitive for all regions.  Users requiring a precise estimate of
completeness in the GLIMPSE~360/Deep GLIMPSE Catalogs should conduct their
own analysis of saturated star density specific to their
region of interest and its impact on Catalog completeness.  We recommend
the  GLIMPSE~360/Deep GLIMPSE Archive in lieu of the Catalog for applications requiring 
high completeness.}
\end{deluxetable}

\begin{deluxetable}{lccccccc}
\tabletypesize{\scriptsize}
\tablecaption{GLIMPSE 360  \& Deep GLIMPSE Archive Completeness1 as a function of band and
magnitude  \label{G360A1.tab}}
\setlength{\tabcolsep}{0.02in}
\tablehead{
\colhead{Background } &
\colhead{11--12} &
\colhead{12--13} &
\colhead{13--14} &
\colhead{14--15} &
\colhead{15--16} &
\colhead{16--17} &
\colhead{17--18} \\
\colhead{(MJy/sr)} &
\colhead{(mag.)} &
\colhead{(mag.)} &
\colhead{(mag.)} &
\colhead{(mag.)} &
\colhead{(mag.)} &
\colhead{(mag.)} &
\colhead{(mag.)} } 
\tablecolumns{8}
\startdata
\cutinheadc{3.6 $\mu$m}
  0.12--  0.21 &  1.00 &  1.00 &  1.00 &  1.00 &  1.00 &  0.87 &  0.13 \\
  0.21--  0.37 &  1.00 &  1.00 &  1.00 &  1.00 &  1.00 &  0.84 &  0.10 \\
  0.37--  0.66 &  1.00 &  1.00 &  1.00 &  1.00 &  0.99 &  0.75 &  0.06 \\
  0.66--  1.17 &  1.00 &  0.99 &  0.99 &  0.98 &  0.96 &  0.60 &  0.03 \\
  1.17--  2.06 &  0.99 &  0.98 &  0.97 &  0.95 &  0.88 &  0.40 &  0.01 \\
  2.06--  3.63 &  0.98 &  0.98 &  0.96 &  0.91 &  0.71 &  0.16 &  0.00 \\
  3.63--  6.41 &  0.96 &  0.93 &  0.85 &  0.71 &  0.33 &  0.03 &  0.00 \\
  6.41-- 11.32 &  0.90 &  0.79 &  0.51 &  0.32 &  0.04 &  0.00 &  0.00 \\
 11.32-- 19.99 &  0.72 &  0.48 &  0.17 &  0.03 &  0.00 &  0.00 &  0.00 \\
 19.99-- 35.29 &  0.27 &  0.09 &  0.02 &  0.00 &  0.00 &  0.00 &  0.00 \\
 35.29-- 62.31 &  0.00 &  0.00 &  0.00 &  0.00 &  0.00 &  0.00 &  0.00 \\
 62.31--110.00 &  0.00 &  0.00 &  0.00 &  0.00 &  0.00 &  0.00 &  0.00 \\
\hline
\cutinheadc{4.5 $\mu$m}
  0.07--  0.12 &  1.00 &  1.00 &  1.00 &  1.00 &  1.00 &  0.88 &  0.13 \\
  0.12--  0.22 &  1.00 &  1.00 &  1.00 &  1.00 &  1.00 &  0.86 &  0.12 \\
  0.22--  0.38 &  1.00 &  1.00 &  1.00 &  1.00 &  1.00 &  0.81 &  0.08 \\
  0.38--  0.66 &  1.00 &  1.00 &  1.00 &  0.99 &  0.98 &  0.70 &  0.05 \\
  0.66--  1.17 &  1.00 &  0.99 &  0.98 &  0.97 &  0.93 &  0.53 &  0.02 \\
  1.17--  2.05 &  0.99 &  0.98 &  0.97 &  0.93 &  0.84 &  0.31 &  0.00 \\
  2.05--  3.60 &  0.98 &  0.97 &  0.94 &  0.89 &  0.61 &  0.09 &  0.00 \\
  3.60--  6.32 &  0.96 &  0.92 &  0.82 &  0.64 &  0.24 &  0.02 &  0.00 \\
  6.32-- 11.09 &  0.89 &  0.80 &  0.50 &  0.32 &  0.03 &  0.00 &  0.00 \\
 11.09-- 19.47 &  0.76 &  0.45 &  0.16 &  0.03 &  0.00 &  0.00 &  0.00 \\
 19.47-- 34.18 &  0.35 &  0.12 &  0.03 &  0.00 &  0.00 &  0.00 &  0.00 \\
 34.18-- 60.00 &  0.06 &  0.03 &  0.00 &  0.00 &  0.00 &  0.00 &  0.00 \\
\hline
\enddata
\end{deluxetable}

\begin{deluxetable}{lccccccc}
\tabletypesize{\scriptsize}
\tablecaption{GLIMPSE 360  \& Deep GLIMPSE  Archive Completeness2 as a function of band and
magnitude  \label{G360A2.tab}}
\setlength{\tabcolsep}{0.02in}
\tablehead{
\colhead{Background } &
\colhead{11--12} &
\colhead{12--13} &
\colhead{13--14} &
\colhead{14--15} &
\colhead{15--16} &
\colhead{16--17} &
\colhead{17--18} \\
\colhead{(MJy/sr)} &
\colhead{(mag.)} &
\colhead{(mag.)} &
\colhead{(mag.)} &
\colhead{(mag.)} &
\colhead{(mag.)} &
\colhead{(mag.)} &
\colhead{(mag.)}
 } 
\tablecolumns{8}
\startdata
\cutinheadc{3.6 $\mu$m}
  0.12--  0.21 &  0.99 &  1.00 &  1.00 &  1.00 &  1.00 &  0.87 &  0.13 \\
  0.21--  0.37 &  0.99 &  1.00 &  1.00 &  1.00 &  1.00 &  0.84 &  0.10 \\
  0.37--  0.66 &  0.99 &  1.00 &  1.00 &  0.99 &  0.99 &  0.74 &  0.05 \\
  0.66--  1.17 &  0.99 &  0.99 &  0.98 &  0.97 &  0.95 &  0.59 &  0.02 \\
  1.17--  2.06 &  0.98 &  0.97 &  0.96 &  0.94 &  0.87 &  0.38 &  0.01 \\
  2.06--  3.63 &  0.97 &  0.97 &  0.95 &  0.91 &  0.70 &  0.15 &  0.00 \\
  3.63--  6.41 &  0.95 &  0.91 &  0.84 &  0.69 &  0.31 &  0.02 &  0.00 \\
  6.41-- 11.32 &  0.87 &  0.78 &  0.50 &  0.29 &  0.04 &  0.00 &  0.00 \\
 11.32-- 19.99 &  0.68 &  0.45 &  0.14 &  0.03 &  0.00 &  0.00 &  0.00 \\
 19.99-- 35.29 &  0.27 &  0.07 &  0.02 &  0.00 &  0.00 &  0.00 &  0.00 \\
 35.29-- 62.31 &  0.00 &  0.00 &  0.00 &  0.00 &  0.00 &  0.00 &  0.00 \\
 62.31--110.00 &  0.00 &  0.00 &  0.00 &  0.00 &  0.00 &  0.00 &  0.00 \\
\hline
\cutinheadc{4.5 $\mu$m}
  0.07--  0.12 &  1.00 &  1.00 &  1.00 &  1.00 &  1.00 &  0.85 &  0.12 \\
  0.12--  0.22 &  1.00 &  1.00 &  1.00 &  1.00 &  1.00 &  0.84 &  0.11 \\
  0.22--  0.38 &  1.00 &  1.00 &  1.00 &  1.00 &  0.99 &  0.78 &  0.07 \\
  0.38--  0.66 &  1.00 &  1.00 &  1.00 &  0.99 &  0.98 &  0.68 &  0.04 \\
  0.66--  1.17 &  1.00 &  0.99 &  0.98 &  0.97 &  0.93 &  0.50 &  0.02 \\
  1.17--  2.05 &  0.99 &  0.98 &  0.97 &  0.93 &  0.83 &  0.29 &  0.00 \\
  2.05--  3.60 &  0.98 &  0.97 &  0.94 &  0.89 &  0.59 &  0.08 &  0.00 \\
  3.60--  6.32 &  0.96 &  0.92 &  0.82 &  0.63 &  0.23 &  0.01 &  0.00 \\
  6.32-- 11.09 &  0.89 &  0.80 &  0.49 &  0.32 &  0.03 &  0.00 &  0.00 \\
 11.09-- 19.47 &  0.76 &  0.43 &  0.16 &  0.03 &  0.00 &  0.00 &  0.00 \\
 19.47-- 34.18 &  0.35 &  0.10 &  0.03 &  0.00 &  0.00 &  0.00 &  0.00 \\
 34.18-- 60.00 &  0.06 &  0.03 &  0.00 &  0.00 &  0.00 &  0.00 &  0.00 \\
\hline
\enddata
\end{deluxetable}

\clearpage

\clearpage

\clearpage

\clearpage

\clearpage

\clearpage

\clearpage

\clearpage

\clearpage

\clearpage

\clearpage

\clearpage

\clearpage

\clearpage

\clearpage

\clearpage

\end{document}